\documentclass[conference]{IEEEtran}
\usepackage{fancyhdr}
\IEEEoverridecommandlockouts
\def\BibTeX{{\rm B\kern-.05em{\sc i\kern-.025em b}\kern-.08em
    T\kern-.1667em\lower.7ex\hbox{E}\kern-.125emX}}

\usepackage{comment}
\usepackage{cite}
\usepackage{multirow}
\usepackage{amsmath,amssymb,amsfonts}
\usepackage{algorithmic}
\usepackage{graphicx}
\usepackage{textcomp}
\usepackage{xcolor}
\usepackage{todonotes}
\usepackage{subcaption}
\usepackage{caption}
\usepackage{todonotes}
\usepackage{algorithmic}
\usepackage[linesnumbered, ruled]{algorithm2e}
\usepackage{lipsum}
\usepackage{url}
\usepackage{authblk}
\usepackage{blindtext}
\usepackage[switch]{lineno}
\usepackage{booktabs}
\usepackage{mathrsfs} 
\usepackage{listings}
\usepackage{balance}
\usepackage{bm}

\usepackage{xspace}
\usepackage[nolist]{acronym}
\usepackage{enumerate}
\usepackage{enumitem}
\usepackage{ifthen}

\newboolean{GuideOn} 
\setboolean{GuideOn}{False}

\newcommand{\shaheen}{\texttt{Shaheen}\xspace}
\newcommand{\leonardo}{\texttt{Leonardo}\xspace}
\newcommand{\summit}{\texttt{Summit}\xspace}

\newcommand{\ibex}{\texttt{Ibex}\xspace}
\newcommand{\alps}{\texttt{Alps}\xspace}
\newcommand{\frontier}{\texttt{Frontier}\xspace}

\newcommand{\parsec}{\texttt{PaRSEC}\xspace}
\newcommand{\starpu}{\texttt{StarPU}\xspace}

\newcommand{\build}{\texttt{Build}\xspace}

\newcommand{\reduce}[1]{}

\begin{document}

\title{Toward 
Capturing 
Genetic Epistasis From\\ 
Multivariate Genome-Wide Association Studies\\
Using 
Mixed-Precision Kernel Ridge Regression}

\author[1,5]{Hatem Ltaief}
\author[1,2,6]{Rabab Alomairy}
\author[3,7]{Qinglei Cao}
\author[1,5]{Jie Ren}
\author[4,8]{Lotfi Slim}
\author[4,9]{Thorsten Kurth}
\author[4,10]{\\Benedikt Dorschner}
\author[1,5]{Salim Bougouffa}
\author[4,11]{Rached Abdelkhalak}
\author[1,5]{David E. Keyes}
\affil[1]{Computer, Electrical and Mathematical Sciences and Engineering Division, \protect \\
King Abdullah University of Science and Technology, KSA.}
\affil[2]{Computer Science \& Artificial Intelligence Laboratory, \protect \\Massachusetts Institute of Technology, USA.}
\affil[3]{Department of Computer Science, Saint Louis University, USA.}
\affil[4]{NVIDIA, USA.}
\affil[5]{\textit {\{Firstname.Lastname\}@kaust.edu.sa}}
\affil[6]{\textit 
{rabab.alomairy@mit.edu}\quad$^{7}$\textit
{qinglei.cao@slu.edu}\quad$^{8}$\textit {lslim@nvidia.com}
}
\affil[9]{\textit 
{tkurth@nvidia.com}\quad$^{10}$\textit 
{bdorschner@nvidia.com}\quad$^{11}$\textit 
{rabdelkhalek@nvidia.com}
}

\maketitle

\thispagestyle{fancy}
\lhead{}
\rhead{}
\chead{}
\lfoot{\footnotesize{
SC24, November 17-22, 2024, Atlanta, GA, USA
}} 
\rfoot{}
\cfoot{}
\renewcommand{\headrulewidth}{0pt} \renewcommand{\footrulewidth}{0pt}

\begin{abstract}
\ifthenelse {\boolean{GuideOn}} {\textbf{(150 word max)}\\}{}

We exploit the widening margin in tensor-core performance between [FP64/FP32/FP16/INT8,FP64/FP32/FP16/FP8/INT8] on NVIDIA [Ampere,Hopper] 
GPUs to boost the performance of output accuracy-preserving mixed-precision computation of Genome-Wide Association Studies (GWAS) of 305K patients from the UK BioBank, the largest-ever GWAS cohort studied for genetic epistasis using a multivariate approach. Tile-centric adaptive-precision linear algebraic techniques motivated by reducing data motion gain enhanced significance with low-precision GPU arithmetic. At the core of Kernel Ridge Regression (KRR) techniques for GWAS lie compute-bound cubic-complexity matrix operations that inhibit scaling to aspirational dimensions of the population, genotypes, and phenotypes. We accelerate KRR matrix generation by redesigning the computation for Euclidean distances to engage INT8 tensor cores while exploiting symmetry. We accelerate solution of the regularized KRR systems by deploying a new four-precision Cholesky-based solver, which, at 1.805 mixed-precision ExaOp/s on a nearly full Alps system, outperforms the state-of-the-art CPU-only REGENIE GWAS software by five orders of magnitude.

\end{abstract}

\vspace{0.3in}

\begin{IEEEkeywords}
Multivariate Genome-wide Association Studies, Kernel Ridge Regression, 
Nonlinear genotype-phenotype relationships, 
UK BioBank data,
Tile-centric matrix computations, Mixed precision, 
Dynamic runtime system, 
GPU accelerators.
\end{IEEEkeywords}

\section{JUSTIFICATION FOR THE GORDON BELL PRIZE}
\ifthenelse {\boolean{GuideOn}} {
\textbf{(50 word max)
indicate what implementation or performance “high watermark” was achieved (rather than the science that was enabled)\\
}
} {
}
High-performance tile-centric matrix computations for Kernel Ridge Regression. End-to-end GWAS software supporting the largest-ever multivariate study of 305K patients from UK BioBank real datasets and 13M patients from synthetic datasets. Application-worthy FP32 accuracy using four precisions, including INT8 and FP8. 
Near-perfect weak-scaling on full-scale \alps, achieving $1.805$ mixed precision ExaOp/s.

\vspace{.7in}
\section{PERFORMANCE ATTRIBUTES}
\begin{center}
\begin{table}[h]
\begin{tabular}{l|l}
\hline
Performance Attributes       & Value                                               \\ \hline
Problem Size & 305K patients [real data, UK BioBank] \\      
                & 300K patients [synthetic data, msprime]\\
                & 13M patients  [synthetic data, random fill]\\
                \hline
Category of achievement      & Scalability,   performance, \\
& time to solution \\ 
\hline
Type of method used          & Kernel Ridge Regression                 \\ 
\hline
Results reported on basis of & Whole GWAS application:                     \\ 
                             & - mixed precision distance computation \\
                             & - mixed precision Cholesky factorization \\
                             & - mixed precision triangular solve \\
                             \hline
Precision reported           & FP64, FP32, FP16, FP8, INT8                \\
\hline
System scale                 & 2/3 of \summit ($18,432$ V100 GPUs)\\
                             & 1/3 of \leonardo ($4,096$ A100 GPUs)\\
                             & full \frontier ($36,100$ MI250X GPUs)\\
                             & 4/5 of \alps ($8,100$ GH200 Superchips)\\
                             & Sustained $1.805$ mixed precision ExaOp/s \\
                             \hline
Measurement mechanism        & Timers, Flops\\
\hline
\end{tabular}
\end{table}
\end{center}

\section{OVERVIEW OF THE PROBLEM}
\label{sec:overview}
\ifthenelse {\boolean{GuideOn}} {
{\bf description of the problem and its importance, in terms understandable to a non-specialist (1 p max)}
} { }

Genome-Wide Association Studies (GWAS) analyze DNA sequence variations spanning an entire genome (human or other) in order
to identify genetic risk factors for 
diseases or other traits within a population. A main goal of GWAS is to use genetic factors to make predictions about individuals at risk and to identify the biological underpinnings of disease. This aids in the development of new diagnostic and therapeutic strategies \cite{Uffelmann2021}.

\begin{figure}[htp]
\center
\includegraphics[width=\linewidth]{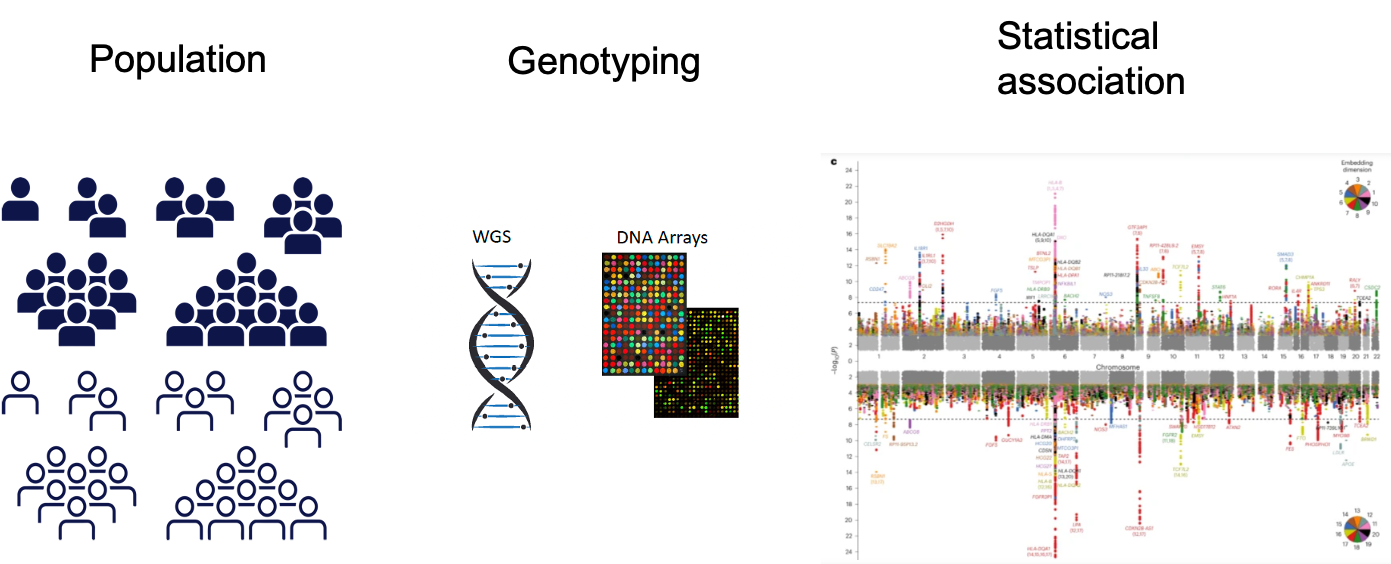}
\caption{Genome-wide association study~\cite{gwasnature}.}
\label{fig:gwas}
\end{figure}

In a typical GWAS workflow, sketched in Fig.~\ref{fig:gwas}, several thousand to several million Single Nucleotide Polymorphisms (SNPs), the standard unit of genetic variation, are genotyped for large cohorts reaching into the millions 
of individuals. Extensive phenotypic information related to various traits or characteristics is also compiled. This can include disease diagnoses, measurements of physical traits (e.g., height, weight), behavioral assessments, and laboratory test results~\cite{Bycroft2018,Kurki2023}. However, no matter how extensive the phenotypic information, it is always orders of magnitude smaller than the accompanying genotypic information in terms of the number of descriptors. This same observation often applies to the number of individuals in the GWAS cohort, which is typically smaller than the number of genotyped SNPs (though not with the largest synthetic cohort employed in this paper). In statistics, this is often referred to as the `large p, small n' problem~\cite{Huynh2020}.

Amidst the multitude of tools available from the high-dimensional statistics toolkit, the dominant approach in GWAS has been univariate statistical testing where each SNP is independently tested for association with the trait of interest, without regard for potential interactions or correlations with other genetic loci (positions). Interactions between distant loci, or epistasis, accounts for the fact that genes do not operate in isolation, but interact with each other in complex ways. This is demonstrated by the polygenic nature of several human traits and diseases where epistasis helps shape the genetic complexity of these traits by modulating their expression and inheritance patterns \cite{Phillips2008}. The strong correlation between neighboring SNPs is known as linkage disequilibrium (LD), and it refers to the non-random association of alleles at different loci along a chromosome. This could result in false positive associations, as the detected associations might be attributed to LD between the identified SNPs and the true causal variants. This comes in addition to violating the independence assumption on which multiple-comparison correction methods rely to control for the inflated risk of so-called ``Type I'' errors, where a significant association is predicted between a genetic variant and the trait or disease, even though there is no actual association. A natural strategy to address both limitations is to employ multivariate approaches, which can model the collective association of a set of SNPs with one or more phenotypes of interest. Multivariate approaches can also help account for potential confounding factors such as population structure or demographic and environmental factors. 

Linear models are typically the initial choice when considering multivariate methods. However, within the context of GWAS characterised by high predictor-to-sample ratio, linear models may suffer from multiple limitations. For instance, overfitting often leads to a diminished generalization performance of the fitted model, undercutting its utility in precision medicine applications. Moreover, the task of variable selection poses challenges in high-dimensional settings, potentially resulting in the detection of spurious associations. Furthermore, from a numerical perspective, the resolution of linear models in such a setting necessitates computing the inverse of an ill-conditioned matrix, namely the sample-wise correlation matrix. Penalized regression approaches \cite{Hastie2009} have been developed to overcome the aforementioned limitations. For example, Ridge Regression (RR) adds to the ordinary least squares objective function a penalty term proportional to the square of the Euclidean norm of the coefficients. The RR method shrinks coefficients towards zero but retains all predictors, unlike LASSO~\cite{Tibshirani1996}, another penalized regression model, which eliminates predictors by setting some coefficients exactly to zero.

The RR approach still models the relationship between predictors and the response variable as linear, potentially overlooking the complexity and nonlinear nature of genotype-phenotype relationships. Nonlinearity in GWAS can arise from factors like epistasis, gene-environment interactions, and non-additive genetic effects. One widely-used approach to address this limitation is through kernel methods \cite{Hofmann2008}, which transform the input data into a higher-dimensional feature space where nonlinear relationships can be more effectively captured and modeled. This makes Kernel Ridge Regression (KRR) an appealing option for GWAS applications. 
\begin{algorithm}[b]
\begin{algorithmic}[1]
\caption{KRR driver for multivariate GWAS.}\label{alg:krr}
    \STATE \textbf{Input}
    \STATE $N_{P1}$: \# of Patients in training set
    \STATE $N_{P2}$: \# of Patients in testing set
    \STATE $N_S$: \# of SNPs
    \STATE $N_{Ph}$: \# of Phenotypes
    \STATE $G$: $N_{P1} \times N_S $ (Training genotype matrix)
    \STATE $P_h$: $N_{P1} \times N_{Ph}$ (Training phenotype matrix)
    \STATE $T$: $N_{P2} \times N_S$ (Testing genotype matrix)
    \STATE $\gamma$: kernel bandwidth
    \STATE $\alpha$: regularization parameter
    \STATE \textbf{Output}
    \STATE $K$: $N_{P1} \times N_{P1}$ (KRR matrix)
    \STATE $W$: $N_{P1} \times N_{Ph}$ (Weight matrix)
    \STATE $P_r$: $N_{P2} \times N_{Ph}$ (Predictions)
    \STATE \textbf{Phase 1:} $\textsc{Build}(\gamma,G,G,K)$
    \STATE \textbf{Phase 2:} $\textsc{Associate}(\alpha,K,P_h,W)$
    \STATE \textbf{Phase 3:} $\textsc{Predict}(\gamma, G,T, W,P_r)$
\end{algorithmic}
\end{algorithm}
Kernel methods are computationally intensive, especially when handling large datasets or complex kernel functions. The overall three-phase KRR workflow is shown in Algorithm~\ref{alg:krr} for GWAS. There are two important hyperparameters that influence the quantitative and qualitative behaviors of KRR. Higher values of $\alpha$ (Alg. 3, line 3) lead to stronger regularization, forcing the model to generalize better to unseen data by penalizing large coefficients. Smaller values allow the model to fit the training data more closely but may lead to overfitting.  The second KRR hyperparameter, $\gamma$ (Alg. 5, line 4), controls the bandwidth of the kernel function. Higher values of $\gamma$ lead to a more localized influence of nearby data points on the prediction.   Both hyperparameters are typically chosen through techniques such as cross-validation.

The computation of the kernel matrix (phase 1) and its subsequent regularized model fitting (phase 2) can be particularly resource-intensive, as described in Algorithms~\ref{alg:build} and~\ref{alg:associate}, respectively. The third phase computes the predictions, as shown in Algorithm~\ref{alg:predict}. Known as the inference step, it may become time consuming if the association is calculated against several phenotypes. The desire to extend these computationally demanding techniques to large populations, to an expanding array of environmental factors beyond the genome (as in eGWAS~\cite{Virolainen2022}), and to other species of plants and animals, some of which possess larger genome sizes than humans and are of economic or environmental significance, highlights the importance of enhancing the efficiency of the underlying linear algebraic techniques that form the foundation of these methods.
\begin{algorithm}
\begin{algorithmic}[1]
\caption{Build the KRR matrix.}\label{alg:build}
    \STATE \textbf{Procedure} \textsc{Build}$(\gamma,G_1,G_2,K)$
    \STATE $N_{P1} \leftarrow$ \text{rowsize($G_1$)}
    \STATE $N_{P2} \leftarrow$ \text{rowsize($G_2$)}
    \STATE $K \leftarrow$ \text{zeros}($N_{P1}, N_{P2}$)
    \FOR{$i$ in range($1,N_{P1}$)}
        \FOR{$j$ in range($1,N_{P2}$)}
            \STATE $K[i, j] \leftarrow \textsc{KernelMatrix}(\text{type}, \gamma, G_1[i,:], G_2[j,:])$
        \ENDFOR
    \ENDFOR
\end{algorithmic}
\end{algorithm}
\begin{algorithm}
\begin{algorithmic}[1]
\caption{Associate genotype-phenotype.}\label{alg:associate}    \STATE \textbf{Procedure} \textsc{Associate}$(\alpha,K,P_h,W)$
    \STATE Factorize the KRR matrix 
    \STATE $\Tilde{K} \leftarrow \textsc{Factorize}(K + \alpha \cdot Id)$    
    \STATE Solve for $W$ 
    \STATE $W \leftarrow \textsc{Solve}(\Tilde{K},P_h)$
\end{algorithmic}
\end{algorithm}
\begin{algorithm}
\begin{algorithmic}[1]
\caption{Predict for a new cohort.}\label{alg:predict} 
    \STATE \textbf{Procedure} \textsc{Predict}($\gamma,G,T,W,P_r$)
    \STATE $N_{P1} \leftarrow$ \text{rowsize($G$)}
    \STATE $N_{P2} \leftarrow$ \text{rowsize($T$)}    
    \STATE $K$: $N_{P2} \times N_{P1}$ (test-training kernel matrix)
    \STATE $\textsc{Build}(\gamma,T, G, K)$
    \STATE $P_r \leftarrow K \times W$
\end{algorithmic}
\end{algorithm}

\section{CURRENT STATE OF THE ART}
\label{sec:soa}
\ifthenelse {\boolean{GuideOn}} {
\textbf{quantitative discussion of current SoA, with emphasis on performance-related aspects  (1 p max)\\
}
} {
}

Linear Mixed Models (LMMs) have emerged as a preferred tool in GWAS due to their ability to accommodate population structure, address relatedness between individuals, and accurately model the effects of all genotyped SNPs. However, early LMM methods posed computational challenges with complexities either of $\mathcal{O}(N_S N_P^2)$ or $\mathcal{O}(N_S^2N_P)$, where $N_P$ represents the size of the GWAS cohort and $N_S$ denotes the number of genotyped SNPs. Given the scale of current biobanks housing millions of SNPs for millions of individuals, LMM analysis can be computationally prohibitive. This motivation has spurred the development of novel LMM approaches that provide improved computational efficiency while minimizing the loss of statistical power. BOLT-LMM ~\cite{Loh2015} exemplifies such an approach, boasting linear complexity in both dimensions, $\mathcal{O}(N_SN_P)$. This improved complexity is attained by sidestepping the computation of the genotype relationship matrix (GRM). The GRM matrix is utilized to adjust for genetic relatedness among individuals within the study cohort, involving the calculation of all pairwise genetic similarities across SNP markers. fastGWA~\cite{Jiang2019} introduces an alternative LMM approach for GWAS, avoiding the need for computing the GRM matrix and its inverse through an approximation strategy. It utilizes a sparse GRM, where non-null coefficients are reserved solely for closely related individuals (e.g., those with a relatedness coefficient $>0.05$). The computational bottleneck is countered in a different way in REGENIE~\cite{Mbatchou2021}, an efficient and scalable approach grounded in a pair of stacked ridge regressions. Instead of relying on the entire genotype matrix to approximate the GRM matrix, REGENIE partitions the genome into contiguous segments to extract a smaller set of representative variables for each segment. These representative variables correspond to the predicted values across a range of regularization parameters. The motivation is to capture the unknown number and size of truly associated genetic markers. 

To compute $p$-values for assessing the statistical significance of association of the tested SNP markers, all the aforementioned methods construct individual Linear Mixed Models (LMMs) for each SNP. In these models, the effect of the tested SNP is depicted as a fixed effect, while the impact of the rest of the genome is included as a random effect, modeled using the Genotype Relationship Matrix (GRM). Alternatively, in the absence of computational constraints, more complex LMM models can be considered, incorporating a larger number of SNPs modeled as random effects to jointly test their phenotypic effects. Approaches such as Wald tests and likelihood ratio tests can be employed for such tasks.
From a biological standpoint, the set of tested SNPs may correspond to various genomic regions, such as a complete gene, a regulatory region, or even the entire genome. This approach enables the utilization of the finer granularity provided by Whole Exome Sequencing (WES) or Whole Genome Sequencing (WGS) in contrast to SNP arrays.  
While not an exact match in terms of the considered genotype-phenotype relationships, there has been a recent increase in the use of large regression models for inference in quantitative genetics. Particularly in the high-dimensional setting of GWAS, regularized (i.e., penalized) methods such as RR, LASSO, and Elastic Net (combining penalties from both RR and LASSO) have been employed. Whittaker et al.~\cite{whittaker2000} were the first to propose RR, often referred to as the L2-regularizer, for performing risk prediction in GWAS. The computation of GWAS with RR can be translated into solving a system of linear equations involving the genotype matrix, which is typically large and dense. This process can be decomposed into Level-3 Basic Linear Algebra Subprograms (BLAS) operations, including matrix-matrix multiplication, triangular solves, and symmetric rank-k updates, which are required for the dense Cholesky solver.

GWAS matrices are typically encoded using a mixture of integer (i.e., SNP data) and real numbers (i.e., covariates and other confounding factors such as age, gender, principal components of the genetic relationship matrix, etc.)~\cite{Uffelmann2021}. 
The computational methodology needs to accommodate such data type heterogeneity in the input data while, for efficiency, avoiding performing operations with unnecessarily
high precision arithmetic. This requires redesign of the algorithm to take full advantage of Mixed-Precision (MxP) optimizations, such as data-motion reduction and leveraging specific hardware features such as tensor cores. An MxP approach in the context
of GWAS when computing similarity metrics and custom correlation coefficients on Titan and Summit~\cite{Joubert-gwas-sc18} was recognized with the Gordon Bell Prize of 2018.

All of the methods discussed thus far are linear. However, as previously noted, genotype-phenotype relationships can exhibit more complexity and nonlinear behavior. In response to this, SKAT (Sequence Kernel Association Test) \cite{Wu2011} has emerged as a popular statistical method utilized in GWAS. SKAT aggregates information across multiple variants within a genomic region and tests for association with the phenotype using a kernel-based approach. It introduced the identical-by-state (IBS) kernel, a kernel function that quantifies the genetic similarity between individuals by counting the number of alleles individuals $p_i$ and $p_j$ share (count denoted by $p_i \sim p_j$) that are identical by state at a given SNP locus over the total number of alleles. 
Thanks to its flexible weighting scheme, SKAT can assign higher weights to rare variants or variants predicted to have larger effects, thereby enhancing overall statistical power. 
The Gaussian kernel~\cite{Morota2014} is also frequently employed for GWAS applications, and its predictive and performance advantages are illustrated herein. Algorithm~\ref{alg:def} presents the definitions of these two classes of kernel matrices.
\begin{algorithm}
\caption{Kernel matrix definitions.}\label{alg:def}
\begin{algorithmic}[1]
\STATE \textbf{Function} \textsc{KernelMatrix}($\text{type}, \gamma, p_1, p_2$)
    \STATE $N_S \leftarrow$ \text{size($p_1$)}
    \IF{$\text{type} == \text{`Gaussian'}$}
        \RETURN $e^{-\gamma \cdot \left\| p_1 - p_2 \right\|^2}$
    \ELSIF{$\text{type} == \text{`IBS'}$}
        \RETURN $\frac{p_1\sim p_2}{N_S}$
    \ENDIF
\end{algorithmic}
\end{algorithm}

\section{INNOVATIONS REALIZED}
\label{sec:innovations}
\ifthenelse {\boolean{GuideOn}} {
\textbf{what the innovations are and how they were achieved (2 pp max)\\
}
} {
}
\subsection{Ridge Regression with Precision-Adaptive 
Computations}
\label{subsec:rr}


Ridge Regression (RR) minimizes 
a loss function that includes a sum of squared regression residuals. 
The loss function of RR includes a positive penalty parameter $\lambda$ multiplied by the sum of squared regression weights, also known as the L2-norm of the weights, as follows:

\begin{equation}
|| Y - X \beta ||^{2} + \lambda || \beta ||^2,
\end{equation} 

\noindent
where $X$ is the matrix of size $N_{P1} \times N_S$, with $N_{P1}$ and $N_S$ the number of patients and SNPs in the training set, respectively.
$X$ contains the patients' genomic and additional encoded details, e.g., SNPs,
sex, geographical information, etc. $Y$ is the matrix of right-hand sides of the system and is of size $N_{P1} \times N_{Ph}$, with $N_{Ph}$ is the number of targeted labels. $Y$ contains quantitative measures of embedding factors (e.g., diseases or anthropometric traits like weight and BMI). $\beta$ is of size $N_S \times N_{Ph}$ and corresponds to the solution of the system mapping the factors and the patients. The regularization parameter $\lambda$ controls the magnitude of $\beta$. It prevents the solution $\beta$ from overfitting, as explained in Section~\ref{sec:overview}, by reducing the variance so that the weights are shrunk toward zero. Thus, the equation has a solution even in the presence of multicollinearity and $N_{P1} \ll N_S$.  

$\beta$ can be found by solving the following linear system:
\begin{equation}
\beta = (X^{T} X + \lambda I )^{-1} X^{T} Y.
\label{eq:beta}
\end{equation} 

\noindent
Studied in~\cite{alomairy2022} in an HPC context, this equation can be solved by successive compute-bound Level-3 BLAS/LAPACK calls, i.e., symmetric rank-k updates (SYRK), Cholesky-based solver (POTRF and POTRS), and matrix-matrix multiplication (GEMM), on a large dense encoded patient/SNP matrix.
The multi-precision nature of the GWAS encoded datasets (e.g., integer and real) allows to effectively use Tensor Cores (TCs) from NVIDIA GPUs that support low precisions. 

\begin{figure} [htb!]
\centering
\includegraphics[width=\linewidth]{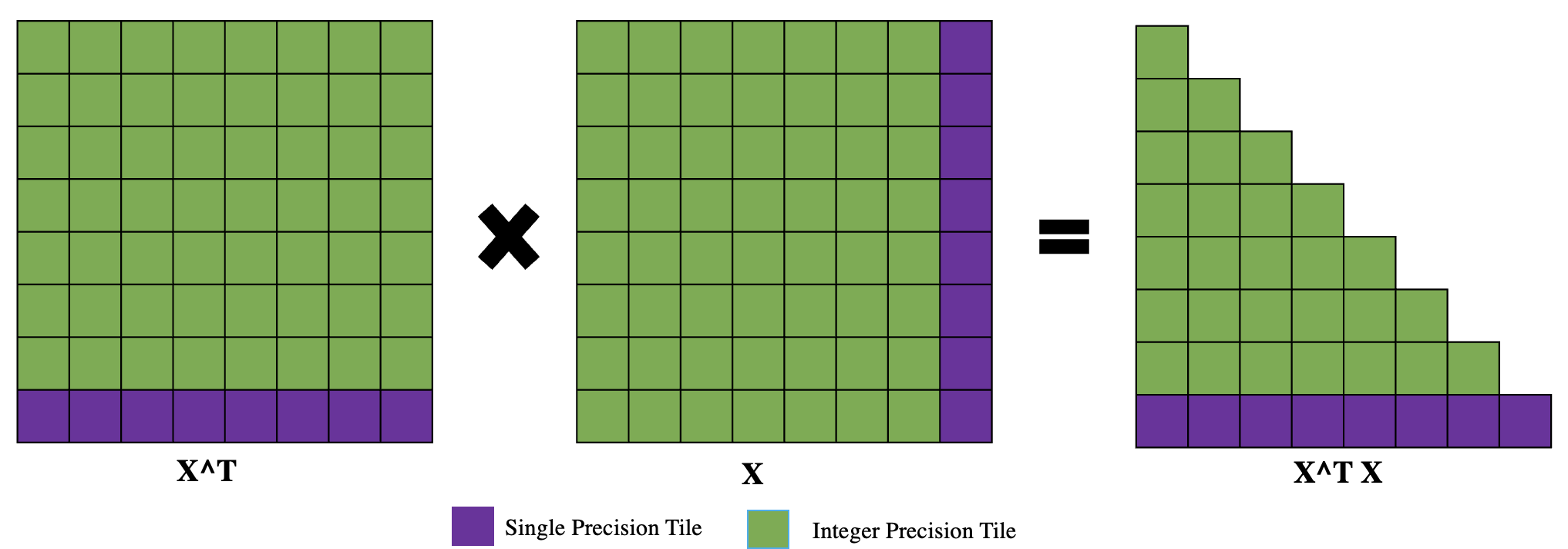}
\caption{Mixed-precision symmetric rank-k update (SYRK).}
\label{fig:mpsyrk}
\end{figure}

The symmetric rank-k update (SYRK) matrix operation is used to compute $X^{T} X$ in  Equation \ref{eq:beta}. 
Each row of $X$ is a vector of SNPs and confounders, which are (generally small) sets of variables beyond the genome whose inclusion is required to prevent spurious SNP-phenotype associations, such as age and geographical and behavioral characteristics. SNPs are represented 
by 0, 1, and 2 which correspond to genetic mutation, and confounders are typically encoded by floating-point numbers, requiring mixed arithmetic. Our 
fine-grained implementation accelerates the SYRK operation by using a mix of integer and floating-point 
TC instructions. The algorithm calls mixed-precision cuBLAS GEMM (i.e., \texttt{cublasGemmEx}) 
for a tile containing integers, and SGEMM (i.e., \texttt{cublasSgemm}) for a tile with floating-point data according to the precision as Fig.~\ref{fig:mpsyrk}. Without fine-grained computations, the few FP32 tiles would contaminate the MxP SYRK operations and render the whole computations in FP32, as restricted by the parallel BLAS and its naming convention. Mixed-precision operations in SYRK use the \texttt{AB8I\_C32I\_OP32I} 
GEMM variant which operates on input operands A/B in INT8, input/output operand C in INT32, and performs accumulation on INT32. The output 
is a symmetric matrix which can be factorized using Cholesky after adding the regularization term $\lambda$ to the diagonal.


The Cholesky factorization is the most time-consuming operation in RR. We redesign and accelerate the Cholesky phase based on half (FP16) and single precision (FP32) floating-point operations. The MxP FP32/FP16 Cholesky factorization
uses a tile-centric adaptive precision~\cite{higham2021mixed,caoGB2022}, orchestrated by a dynamic runtime system (e.g., \starpu~\cite{augonnet2011starpu} and \parsec~\cite{bosilca2013parsec}). The operation $X^{T} Y$ can be computed using matrix-matrix multiplication in single precision because the target matrix $Y$ is usually small and does not benefit from TCs. The Cholesky solve is then performed using forward and backward substitution operations in the full FP32 precision.  


\subsection{Kernel Ridge Regression}
\label{subsec:krr}

\subsubsection{
Engaging INT8 Tensor Cores for Distance Calculations}
\label{subsec:distances}
A key step in the construction of the kernel matrix for KRR is the computation of the Euclidean distance in the genotype dimension between every pair of patients in the $N_S\times N_{P1}$ GWAS matrix $G$, where $N_{P1}$ is the number of patients in the training set.  
The desired output is the strict lower triangle of a symmetric $N_{P1}\times N_{P_1}$ matrix $D$ with elements 
$d_{ij}=\sum_{k=1,NS}(g_{ki}-g_{kj})^2$ for patients $1\le i < j \le N_{P1}$.  Following previous work on redesigning radius calculations for the sphere decoder in wireless communication with general matrix-matrix multiplication (GEMM) for GPU accelerations~\cite{wireless} and more recently~\cite{distance-tc} on computing distances for a general non-symmetric operator using INT8 Tensor Cores (TCs), we exploit the integer encoding of SNPs to compute the squared Euclidean distance for our kernel symmetric matrix using INT8 TCs after unrolling and instruction reordering. To illustrate the trick, consider three patients, $a$, $b$, and $c$, each with two genotypic markers, represented in
$$
G= \left( 
\begin{array}{ccc} 
a_1 &  b_1 & c_1 \\ a_2 & b_2 & c_2
\end{array} 
\right)
.$$
We initialize the integer matrix $D$ with the squared Euclidean norms of the patients down its columns:
$$
D= \left( 
\begin{array}{ccc} 
a_1^2 + a_2^2 & b_1^2 + b_2^2 & c_1^2 + c_2^2 \\
a_1^2 + a_2^2 & b_1^2 + b_2^2 & c_1^2 + c_2^2 \\
a_1^2 + a_2^2 & b_1^2 + b_2^2 & c_1^2 + c_2^2 
\end{array} 
\right).
$$
We form the symmetric $\Tilde{D}\leftarrow D+D^T$ and can then accumulate using the Level-3 BLAS operation SYRK (i.e., symmetric rank-k updates) as follows:
$$\Tilde{D}\leftarrow \Tilde{D}-2* GG^T,$$
leaving (noting that, e.g, $a^2-2ab+b^2=||a-b||^2)$
$$\Tilde{D}=\left( 
\begin{array}{ccc} 
0 & ||b-a||^2 & ||c-a||^2 \\
||a-b||^2 & 0 & ||c-b||^2 \\
||a-c||^2 & ||b-c||^2 & 0
\end{array} 
\right).
$$
Square distances between confounder real variables can be simultaneously accumulated in floating point using BLAS3 SYRK operations in a separate buffer and eventually added to the integer squared distances prior to the exponentiation in the Gaussian kernel in forming the $N_{P1}\times N_{P1}$ kernel matrix $K$ (as seen in Algorithm~\ref{alg:def}). 

\subsubsection{Deploying a Four-Precision Cholesky-Based Solver}
\label{subsubsec:adaptive}
Accelerating matrix computations using AI-centric
low-precision hardware features is a strategic trend in scientific computing~\cite{ltaief2023ISC}. Leveraging tensor core hardware technology not only accelerates floating-point arithmetic but also reduces memory footprint and volume of data communication. Introduced 76 years ago~\cite{Wilkinson-mp}, then revisited in the context of solving a system of linear equations~\cite{Moler-mp} and eigensolvers~\cite{Dongarra-mp}, mixed-precision techniques have been made popular with iterative refinement, especially with the advent of hardware accelerators~\cite{Buttari-iterref,carson2018accelerating,Haidar-iterref}. While this approach recovers all digits of accuracy even for ill-conditioned matrices, it exhibits a large cost in terms of memory footprint due to the requirement to store the matrix operator for each precision involved in the procedure. Recent works revisit the concept of mixed precisions for linear solvers with the idea of avoiding oversolving by computing only up the level of accuracy required by the applications~\cite{doucet-mixedprec,abdulah2018exageostat,caoGB2022,ltaief-reckless,cao-cluster}. With fine-grained task-based matrix computations, we can apply tile-centric adaptive precision~\cite{higham2021mixed} resulting in a tiled mosaic of precisions embedded in a single stored copy of the matrix. With the introduction of new IEEE-compliant FP8 format, we have extended our tile-based Cholesky solver to support now four precisions, i.e., FP64, FP32, FP16, and FP8. To meet the output accuracy expectations of the examples herein, we do not require FP64.

\subsubsection{Studying the Largest-Ever Cohort for Multivariate GWAS}
We evaluate the impact of RR and KRR-based mixed-precision GWAS on a subset of UK BioBank dataset with $305,880$ patients and $43,333$ SNPs. To our knowledge, this is the largest-ever cohort studied for multivariate KRR-based GWAS. The computational challenges due to high arithmetic complexity and lack of flexible solvers for dealing with multi-precision datasets have delayed the wide adoption of multivariate GWAS. Moreover, privacy-considerations associated with such datasets, especially human samples, make it difficult to explore multivariate GWAS on open systems. An interesting byproduct of KRR is that mapping the input data ($G$) into a higher-dimensional space transforms the original integer-encoded dataset into real numbers representing correlations based on computed Euclidean distances, as explained in Section~\ref{subsec:distances}. The nonlinear transformations involved in the kernel matrix cannot be reverse-engineered, allowing the resulting matrix $K$ to be transferred to remote systems without confidentiality concerns. Another interesting feature of direct linear solvers is that one can reuse the factors of $K$ with multiple phenotypes. This is a clear advantage 
over deep learning methods that need to retrain for each phenotype.

\section{HOW PERFORMANCE WAS MEASURED}
\label{sec:measured}
\ifthenelse {\boolean{GuideOn}} {
\emph{(Note that preference is given to performance actually measured [not projected],
based on the entire application [including I/O] and with uniform precision.
Explain in detail if any portion of total runtime was not included in the
measurements, if and where different precisions were used, or any attributes
listed in Section 3 as “other”).\\
what application(s) was used to measure performance (1 p max)\\
system and environment where performance was measured (1 p max)}
} {
}

\subsection{Real / Synthetic Datasets from UK BioBank / Msprime}

We have access to a real clinical dataset obtained from UK BioBank of about half million patients and five million SNPs for each patient. This dataset was anonymized and the research passed ethical review by the UK BioBank organization.
Participants aged 40–69 years were recruited between 2006 and 2010. In this study, we map human genotypes and clinical histories to phenotypes. We extracted several subsets from this dataset to conduct numerical accuracy study and performance experiments. We conduct phenotypic predictions on five common diseases: asthma, allergic rhinitis, depression, general hypertension, and osteoarthritis~\cite{gwasnature}. To further demonstrate the software capabilities on open data, we employ the population genetics software msprime \cite{baumdicker2022efficient} as well as randomly generated datasets up to 13M patients. These large-scale demonstrations herald new opportunities for understanding traits of national-scale human populations or massive assemblies of agricultural accessions.

\begin{figure*} [htb!]
    \centering 
    \includegraphics[width=.3\linewidth]{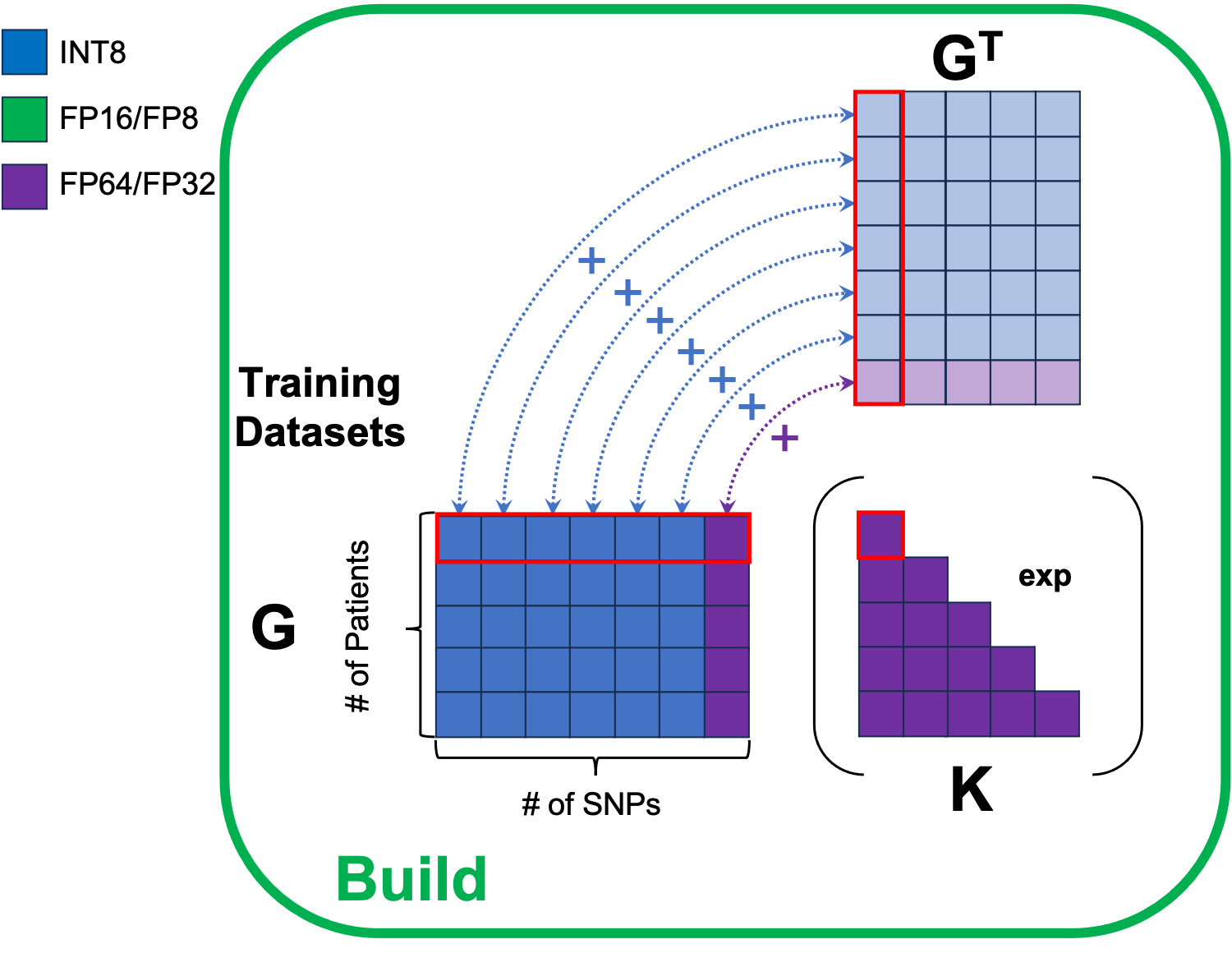}
    \includegraphics[width=.27\linewidth]{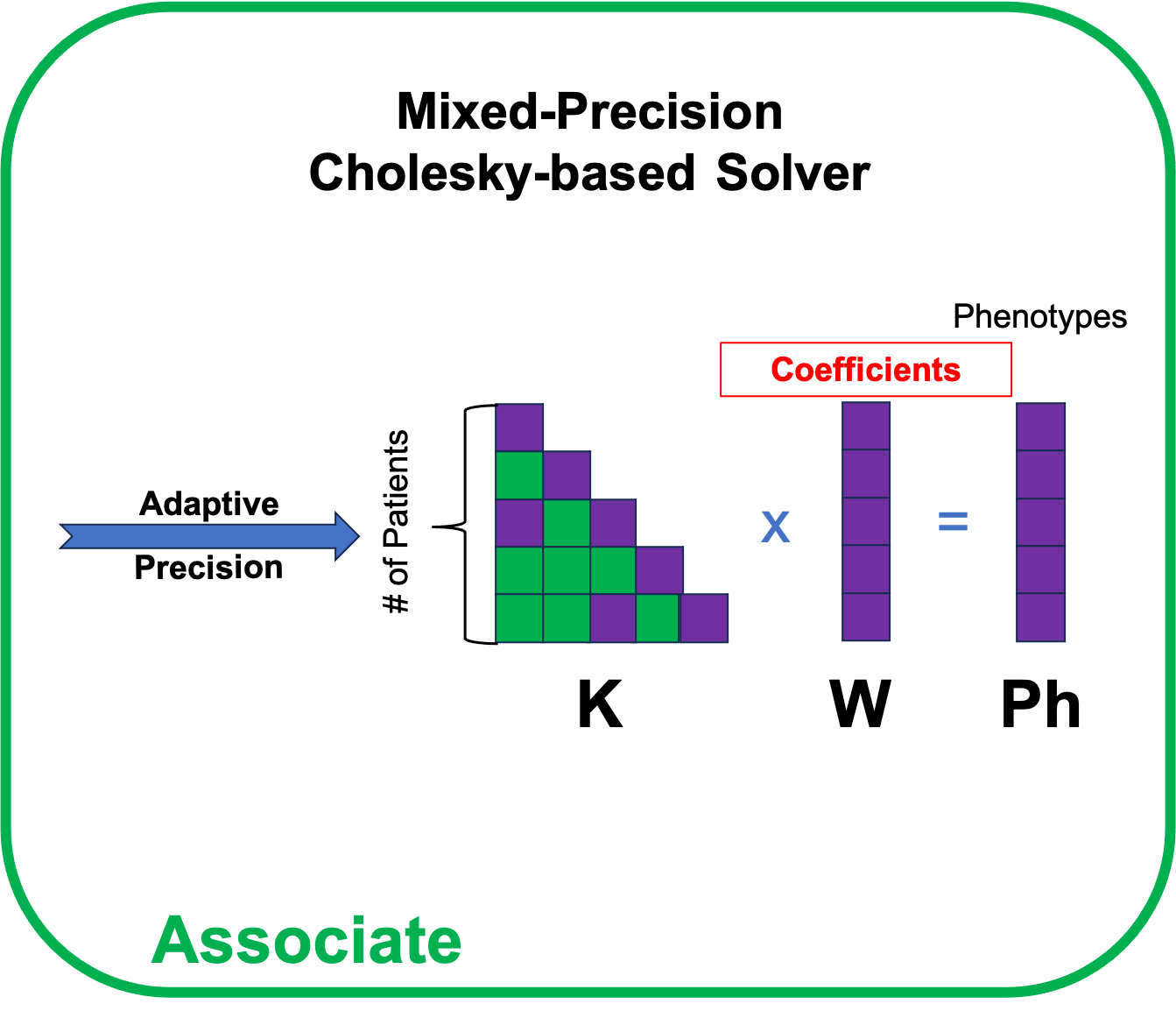}
    \includegraphics[width=.315\linewidth]{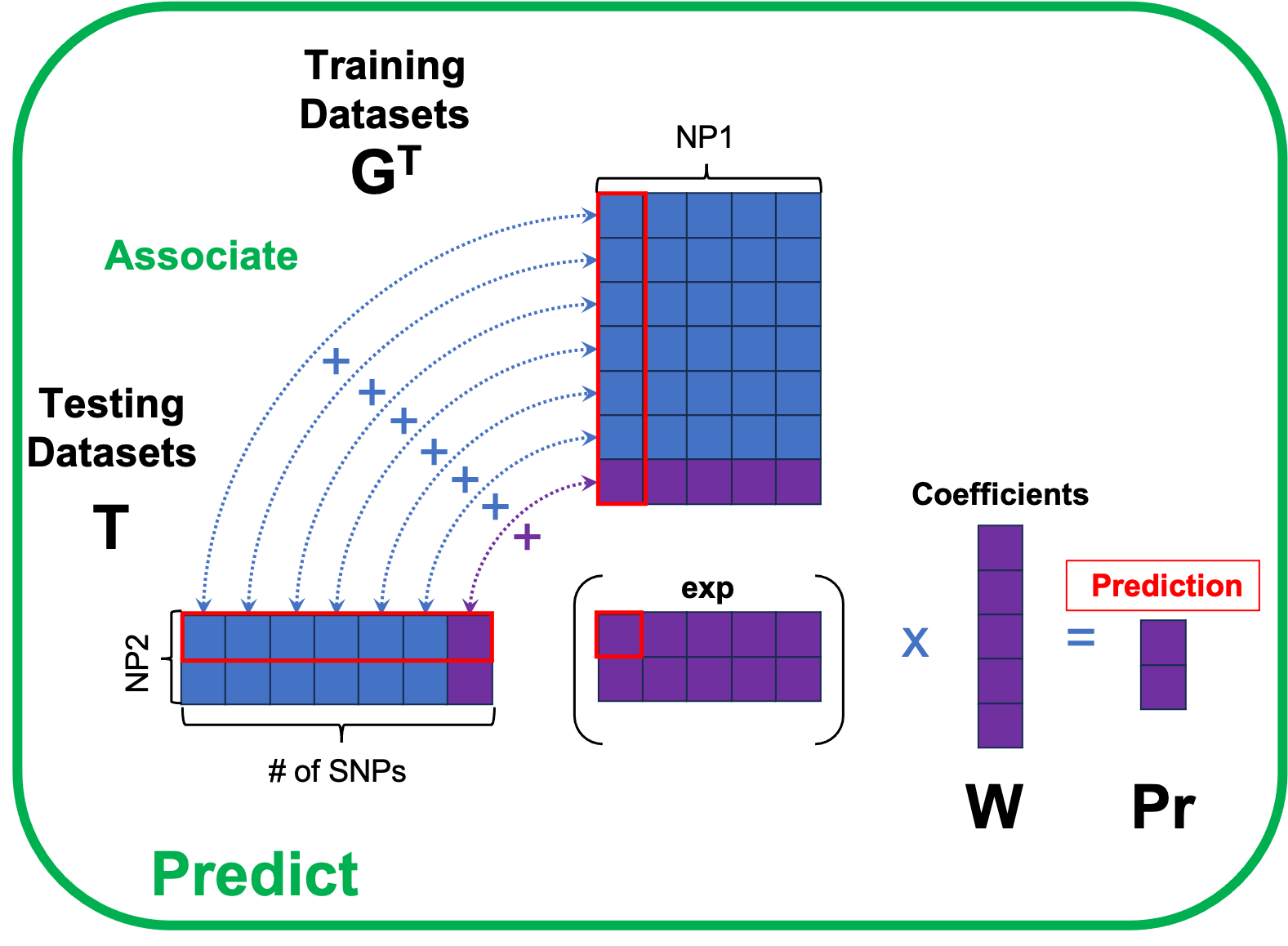}
    \caption{Leveraging the INT8 / FP8 / FP16 / FP32 / FP64 KRR-based multivariate GWAS for genetic epistasis.}
\label{fig:workflow}
\end{figure*}

\subsection{Implementation Details}

\subsubsection{PaRSEC: The Master of Ceremonies}

Our RR and KRR-based multivariate GWAS software is developed in C and relies on the US ECP project~\parsec~\cite{bosilca2013parsec} as the dynamic runtime system. \parsec runs above MPI to initiate inter-node point-to-point and collective communications and uses CUDA/Pthreads within the computational nodes equipped with hardware accelerators. \parsec transparently handles data movement using a dataflow paradigm by traversing a directed acyclic graph where nodes and edges represent tasks and data dependencies, respectively. This asynchronous fine-grained execution permits to overlap communication with computation. Moreover, as part of our mixed-precision endeavor, we empower \parsec with more duties than \textit{simply} orchestrating task scheduling, by using on-the-fly adaptive precisions for datatype conversion. Before taking the decision of moving data from a source to a destination processor, \parsec checks the required precision needed to execute the task at destination and depending on the current precision of the data at the source processors, decides if the conversion should occur at the sender (in case lower precision is needed at the receiver) or at the received side (in case higher precision is needed at the receiver) to save moving unnecessary bytes.

\subsubsection{TC-based Distance Computations}
Casting the basic operations for computing the square of the Euclidean distance into a matrix form to leverage tensor cores may be considered as one of the most innovative techniques for GWAS from an HPC perspective as it will have a significant impact for all GWAS variants [univariate, multivariate / linear, nonlinear]. This technique converts instruction-bound workloads for distance or similarity computations into compute-bound matrix-matrix computations (GEMM). We extend the TC-based distance computations of~\cite{distance-tc} to support the symmetric covariance matrix $K$ that drives our KRR algorithm. We further improve it by realizing that no extra temporary matrices are needed to store intermediate data, but instead the entries can be calculated on-the-fly to save memory footprint. First, one can fold the column entries of the matrix $D$ since they are the same and store $D$ as a one-dimensional-vector instead. Thanks to fine-grained tile-based computations, we can generate $\Tilde{D}$ in tiles on-demand without having to generate it as a whole in memory at any given point in time. We segregate integer from FP32 arithmetic during the accumulation of the symmetric rank-k updates to engage TC and perform the final reduction on the corresponding tile of the $K$ matrix. Before the tile is released, we exponentiate each entry of the tile and this step concludes the \texttt{Build} phase as shown in the left side of Fig.~\ref{fig:gwas}. Fusing these tile-based computational steps with various precisions enables to maximize performance and reduce memory footprint.

\subsubsection{Support for FP8}
The IEEE-compliant FP format comes with two formulations and due to the current definition of the  \texttt{cublasLtMatmul()} API, we can only use the formulation \texttt{CUDA\_R\_8F\_E4M3} to ensure that the operands $A$ and $B$ are of the same type. The main limitation comes from the \texttt{TN} parameters for GEMM, 
which appear only when the Cholesky factorization is performed on the upper triangle.  We address this for the lower triangle by requesting \parsec
to transpose the files in addition to converting them before sending them to their destinations.
Once a tile reaches its destination for the GEMM computation, the \texttt{cublasLtMatmul()} function with FP8 support transposes it back for correct numerical operation, while satisfying the rule for TC FP8 activation.

\subsection{The Overall Multi-Precision GWAS Workflow}
\label{subsec:complexity}
Figure~\ref{fig:workflow} diagrams the \texttt{Build}, 
\texttt{Associate}, and \texttt{Predict} phases of the multivariate GWAS for genetic epistasis, as introduced in Algorithms~\ref{alg:build},~\ref{alg:associate} and~\ref{alg:predict}, respectively. Each phase leverages mixed-precision computations with fine-grained computations and maps each tile-based matrix operation on the corresponding datatype. The \texttt{Build} phase employs kernel fusion and mixed-precision accumulations during the distance calculations performed in INT8 and FP32 (or FP64, if needed). The \texttt{Associate} phase starts with adaptive precision to lower the precision of the off-diagonal tiles depending on the accuracy threshold, as explained in~\cite{caoGB2022}, followed by the mixed-precision Cholesky-based solver to calculate the weight matrix $W$ using FP32 (or FP64), FP16, and FP8. The \texttt{Predict} phase first calls the \texttt{Associate} phase to create the KRR matrix containing the correlations between the training and testing datasets, from which inference can be applied by multiplying the testing dataset with $W$ using FP32 (or FP64). The bulk of the computations for the multivariate GWAS are done in the \texttt{Build} and \texttt{Associate} phases, where MxP SYRK and Cholesky matrix computations account for most operations with an algorithmic complexity of $N_P^2\times N_S$ and $1/3\times N_P^3$, respectively. This overall KRR-based GWAS workflow demonstrates the importance of transforming the irregular memory-bound state-of-the-art GWAS computations into matrix algebra to ride the hardware innovations driven by AI~\cite{ltaief2023ISC}, while providing the necessary HPC framework to tackle complex genetic epistasis.

\section{PERFORMANCE RESULTS }
\label{sec:performance}

\ifthenelse {\boolean{GuideOn}} {
\textbf{include scalability (weak and strong), time to solution, efficiency (of bottleneck resources),
and peak performance (2 pp max)\\}
} {
}

We report precision heatmaps for $K$ and compare the prediction accuracy of KRR-based multivariate GWAS to RR using the same training and testing subsets from the UK BioBank. We then demonstrate the FP8 capabilities of our software solution on synthetic datasets generated by msprime~\cite{baumdicker2022efficient} instead, due to the on-site license restrictions of UK BioBank. We finally show performance benchmarking results on \summit, \leonardo, \frontier, and \alps.

\subsection{Precision Heatmaps}
\label{fig:heatmaps}
\ifthenelse {\boolean{GuideOn}}{\todo[]{Jie}}{}

\begin{figure} [htb!]
  \centering 
  \begin{subfigure}{0.49\linewidth}
    \centering 
    \includegraphics[width=\linewidth]{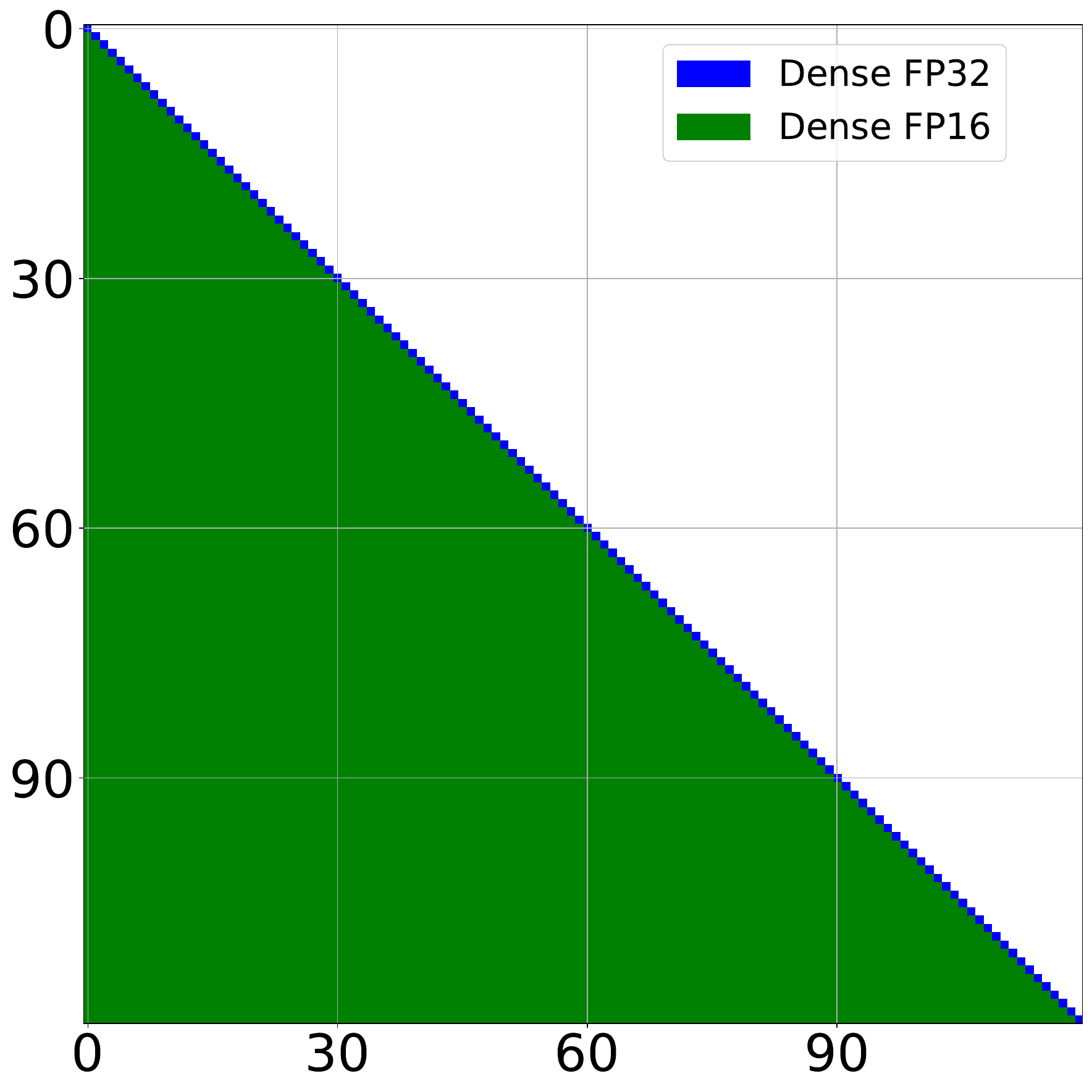}
    \caption{Activating FP16 with A100.}\label{fig:heatmap-fp16}
  \end{subfigure}
    \begin{subfigure}{.49\linewidth}
    \centering 
    \includegraphics[width=\linewidth]{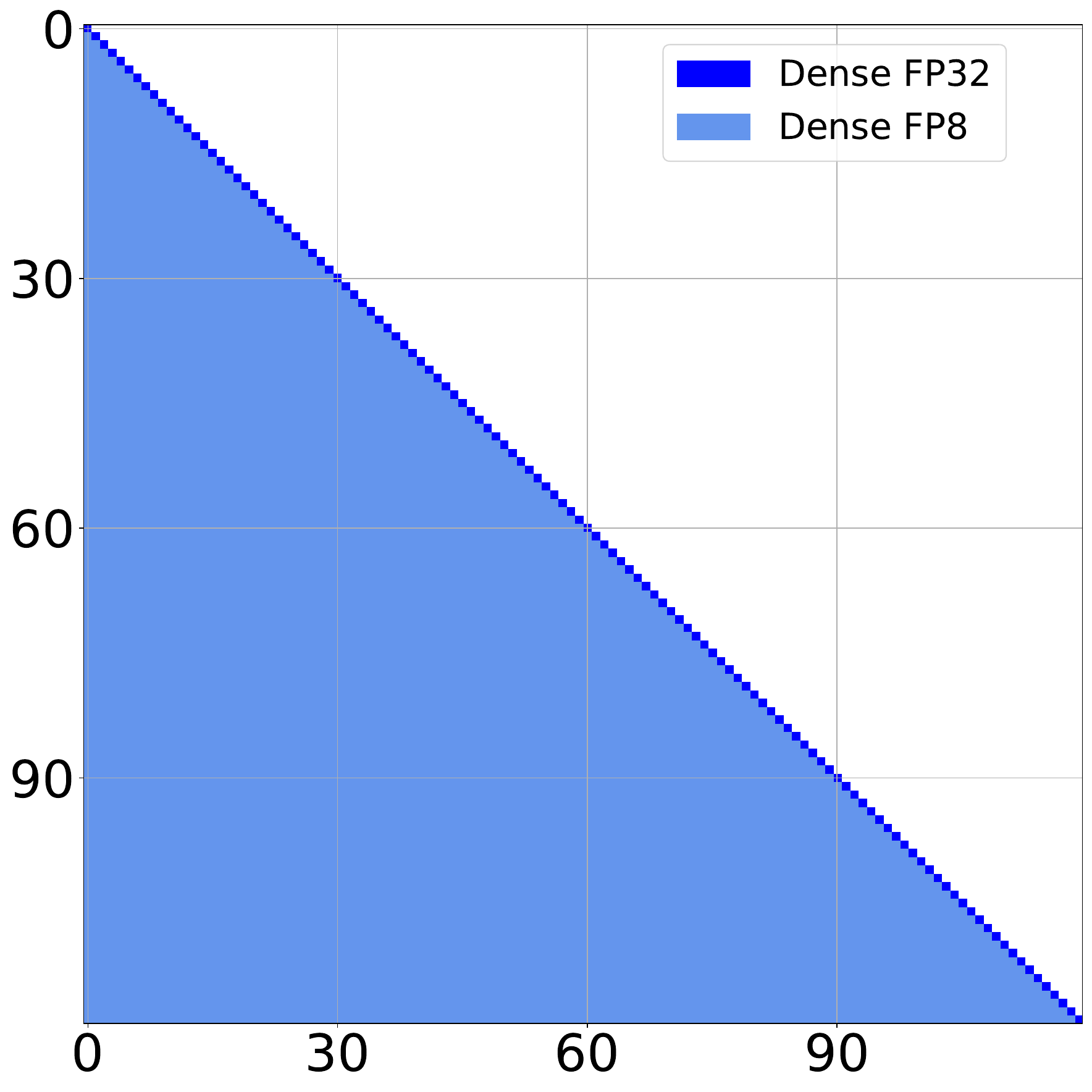}
    \caption{Activating FP8 with GH200.}\label{fig:heatmap-fp8}
  \end{subfigure}
    \caption{Precision heatmaps.}
\label{fig:heatmap}
\end{figure}

We compute the KRR matrix for multivariate GWAS from the \texttt{Build} phase and apply on it the tile-centric adaptive precision technique, as described in Section~\ref{subsubsec:adaptive}.
Figure~\ref{fig:heatmap} shows the precision heatmaps generated at the beginning of the \texttt{Associate} phase -- there are typically no high precision tiles required beyond the diagonal. Higher precision tiles indicate that some individuals in their index range have stronger correlations resulting from the existence of many similar SNPs. Depending on the GPU backends, the adaptive precision technique will adjust the precision of each tile using the lowest numerically appropriate supported datatype, i.e., FP16 and FP8 for A100 and GH200 Superchips, respectively.


\subsection{RR vs. KRR-based Multivariate GWAS}
\label{sec:accuracygwas}
We evaluate the prediction accuracy and robustness of our mixed precision implementation on a subset of the UK BioBank dataset of size $305,880$ patients and $43,333$ SNPs. 80\% of the data is used for training and 20\% is withheld for testing. We test the prediction of five common diseases in the dataset, i.e., \texttt{Hypertension}, \texttt{Asthma}, \texttt{Osteoarthritis}, \texttt{Allergic Rhinitis},  and \texttt{Depression}. 

\subsubsection{Mean Square Prediction Error (MSPE)}
We first evaluate the prediction using the Mean Square Prediction Error (MSPE) relative to the ground truth phenotypes of the testing dataset. MSPE measures the average squared difference between the ground truth and predicted values as follows: 

\begin{equation}
MSPE = \frac{1}{N_{P2}} \sum_{i=1}^{N_{P2}} (Y_i - \hat{Y_i})^2,
\end{equation} 
where $N_{P2}$ is the number of rows (i.e., patients) of the testing dataset, $Y_i$ represents the observed value, and $\hat{Y_i}$ represents inferred values from either RR or KRR.

\begin{figure*}[hbt!]
  \centering 
  \begin{subfigure}{0.32\linewidth}
    \centering 
    \includegraphics[width=.95\linewidth]{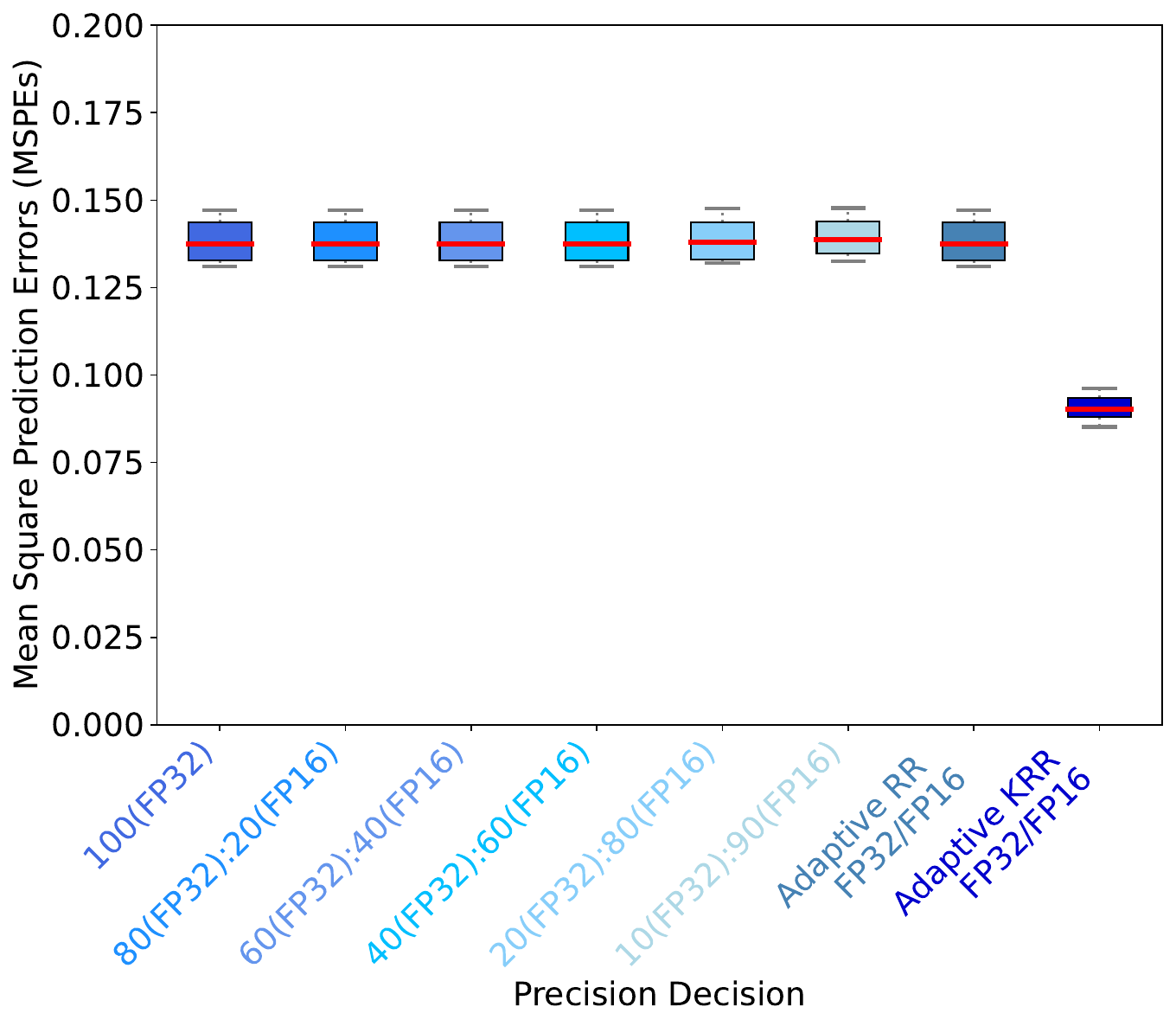}
        \caption{RR vs KRR for Hypertension.}
      \label{fig:msehy_rrVskrr}
  \end{subfigure}
  \begin{subfigure}{0.32\linewidth}
    \centering 
    \includegraphics[width=.95\linewidth]{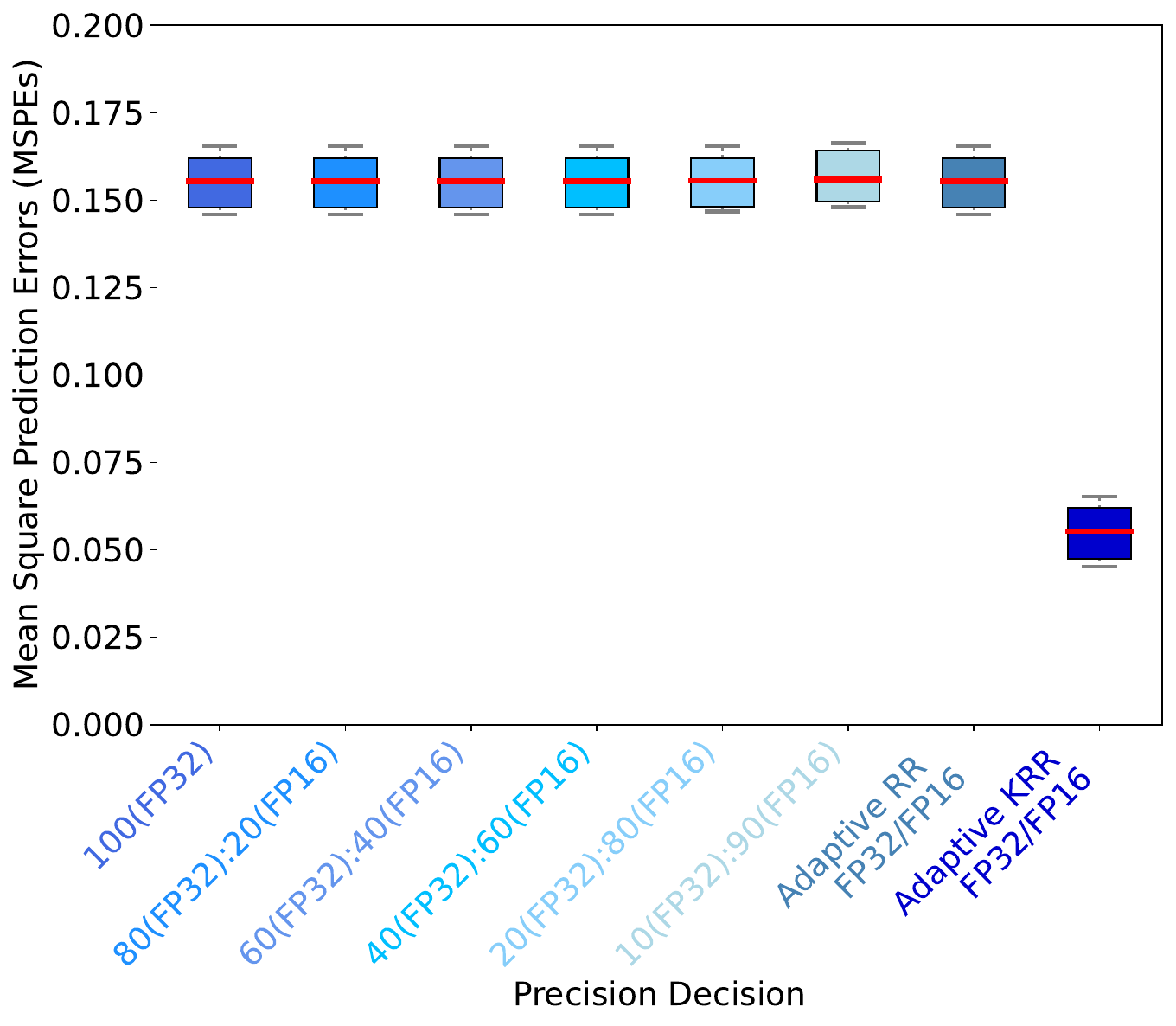}
    \caption{RR vs KRR for Asthma.}
      \label{fig:mseas_rrVskrr}    
  \end{subfigure}
\begin{subfigure}{0.32\linewidth}
    \centering 
    \includegraphics[width=.95\linewidth]{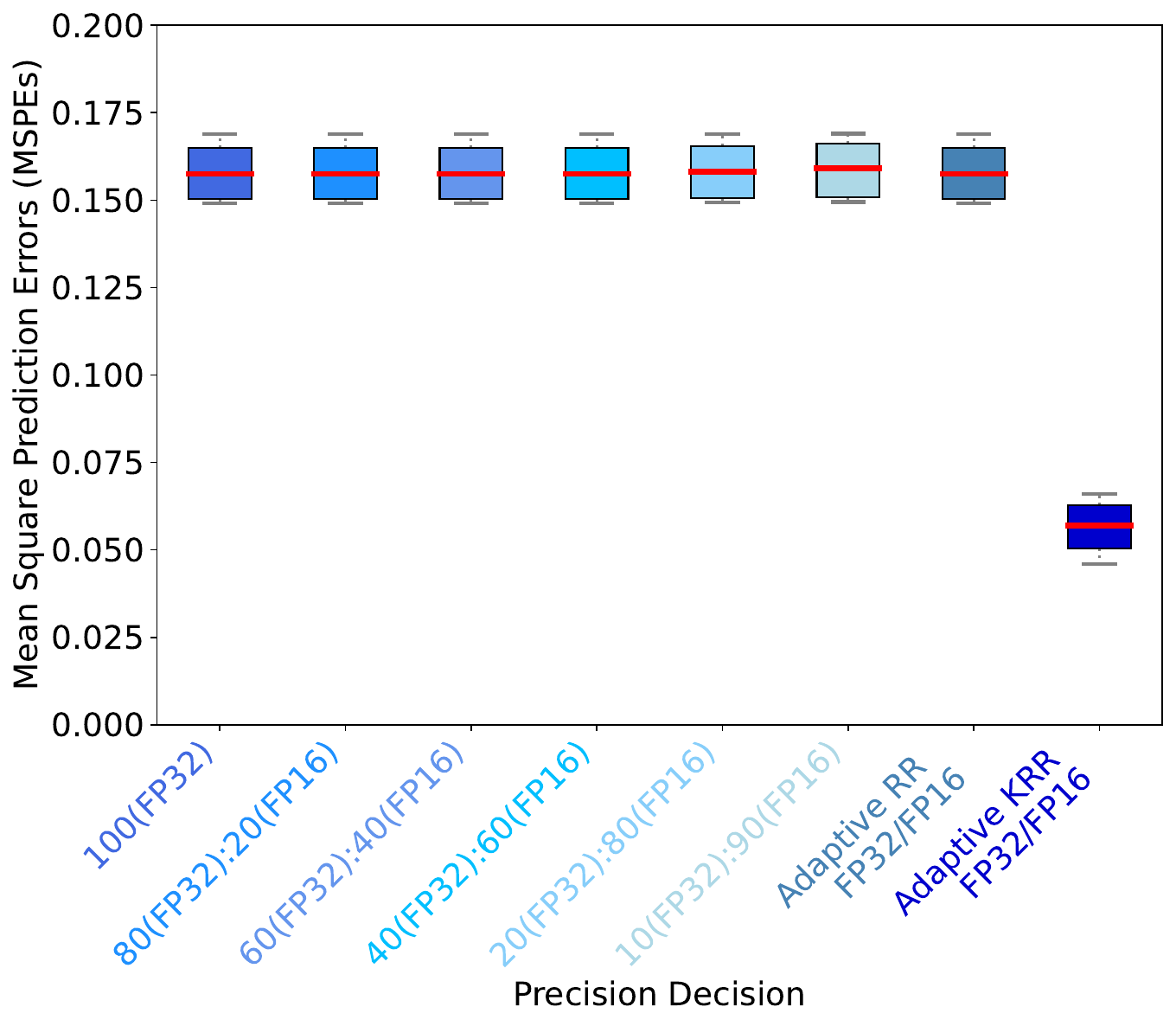}
      \caption{RR vs KRR  for Osteoarthritis.}
      \label{fig:mseost_rrVskrr}
      \end{subfigure}  
     \caption{MSPE comparisons between diseases using $305,880$ patients and $43,333$ SNPs from UK BioBank.}
\end{figure*}

Figures~\ref{fig:msehy_rrVskrr},~\ref{fig:mseas_rrVskrr},
and~\ref{fig:mseost_rrVskrr} 
compare the MSPE of FP32 RR-based multivariate GWAS against the mixed precision approach based on the hand-tuned band technique that applies a given precision on each band of the off-diagonal tiles, creating a rainbow pattern, as introduced in~\cite{abdulah2021accelerating}. The size of each band needs to be determined empirically and can therefore be time-consuming. We report on 100\%, 80\%, 60\%, 40\%, 20\%, and 10\% FP32 (in the first six successively lighter blue boxes), while computing the remaining off-diagonal tiles in FP16. The most constricted band configuration with 10\% FP32 and 90\% FP16 shows a deterioration of MSPE while the previous configurations succeed. In the seventh box in each set,  we identify the required precision for each tile according to the systematic means of~\cite{higham2021mixed}, resulting in FP16 arithmetic for all off-diagonal tiles (see Fig.~\ref{fig:heatmap-fp16}), while achieving the same MSPE as 100\% FP32 for the RR-based multivariate GWAS. In the last box of each disease set, in dark blue, we show the MSPE achieved by KRR-based multivariate GWAS with $\gamma=0.01$ using the tile-centric adaptive precision on the same dataset as RR. KRR achieves a significant prediction superiority for all five diseases (three shown here).

\begin{figure}[t]
\center
\includegraphics[width=0.73\linewidth]{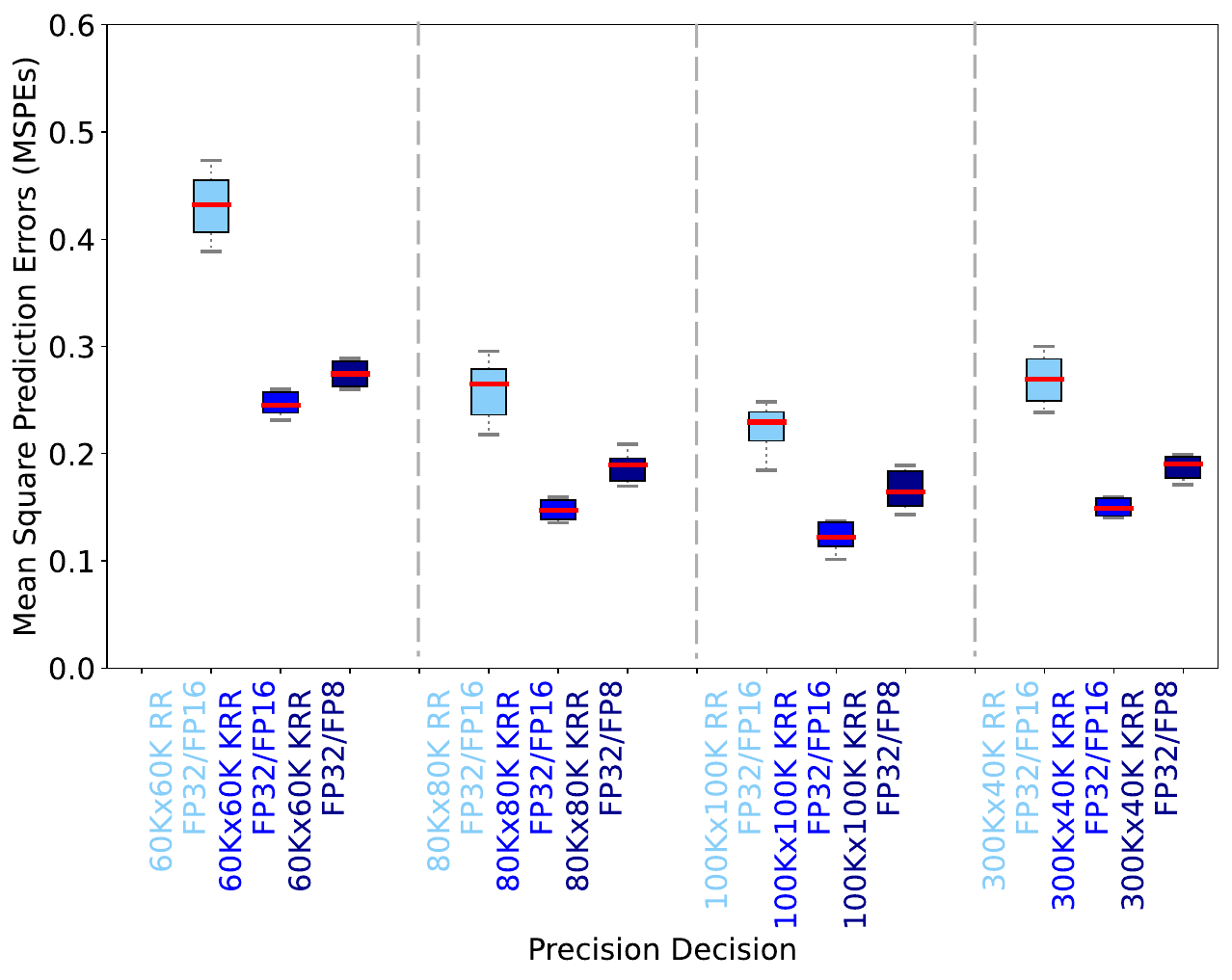}
\caption{MSPE using various $N_P=N_S$,  and $N_P=300$K, $N_S=40$K from msprime~\cite{baumdicker2022efficient}.}
\label{fig:mse_syn}
\end{figure}

Figure~\ref{fig:mse_syn} assesses the KRR-based multivariate GWAS on \alps with GH200 Superchips that hosts FP8 GPU TC hardware features. Since the newly deployed \alps supercomputer does not have a license to host UK BioBank data, we use 300K patients and 40K SNPs generated from msprime~\cite{baumdicker2022efficient} instead, resulting in FP8 arithmetic for all off-diagonal tiles (see Fig.~\ref{fig:heatmap-fp8}) after applying the tile-centric adaptive precision. The MSPE of FP8-enabled KRR is slightly higher than the FP16-enabled KRR but remains lower than FP16-enabled RR. Domain scientists will identify opportunities with best
accuracy/performance trade-offs for phenotypes under study with
the option of FP8 capabilities.  


\subsubsection{Pearson Correlations}
To further emphasize the advantage of KRR over RR-based multivariate GWAS, we assess the Pearson Correlation $\rho$ between two data sets $Y$ and $\hat{Y}$, which is their normalized covariance, 
$-1\le\rho_{Y,\hat{Y}}\le 1$, defined as 
$$\rho_{Y,\hat{Y}}\equiv\frac{cov(Y,\hat{Y})}{\sigma_Y\sigma_{\hat{Y}}},$$
where $\sigma_i$ is the standard deviation (or square-root of the variance) of set $i$.
Table~\ref{tab:pearson} shows the Pearson Correlations between the ground truth
phenotypes of the testing dataset of the UK BioBank patients and, respectively, the predictions
under the RR and the KRR models. Comparing the second and third columns, for the same FP16 precision, the KRR results are more highly correlated with the ground truth than RR (up to four times more), thus demonstrating its capability to model more complex phenotype relationships. The last column (for the synthetic msprime dataset only due to the UK BioBank license restriction) shows a degradation from dropping the lower precision to FP8 in KRR.  The result is still superior to RR in higher precision.\\
\begin{table}[htp!]
\centering
\caption{Pearson correlations: RR vs. KRR.}
\label{tab:pearson}
\begin{tabular}{l|c|c|c}
\hline
\textbf{Phenotypes} & \textbf{RR-FP16}  & \textbf{KRR-FP16} & \textbf{KRR-FP8}\\\hline\hline
\texttt{Hypertension} & 0.2983 &  0.8071 & N/A\\\hline
\texttt{Asthma} & 0.2517 &  0.8205 & N/A\\\hline
\texttt{Osteoarthritis} & 0.3189 &  0.8386 & N/A\\\hline
\texttt{Allergic Rhinitis} &  0.2008 & 0.8652 & N/A\\\hline
\texttt{Depression} & 0.2041 &  0.8454 & N/A\\\hline
\texttt{Synthetic [msprime]} & 0.3418 &  0.6989 & 0.5633 \\\hline
\end{tabular}
\end{table}

\subsection{Performance of TC-based Distance Calculations (\texttt{Build})}

\begin{figure}[htb!]  
    \centering 
    \includegraphics[width=0.8\linewidth]{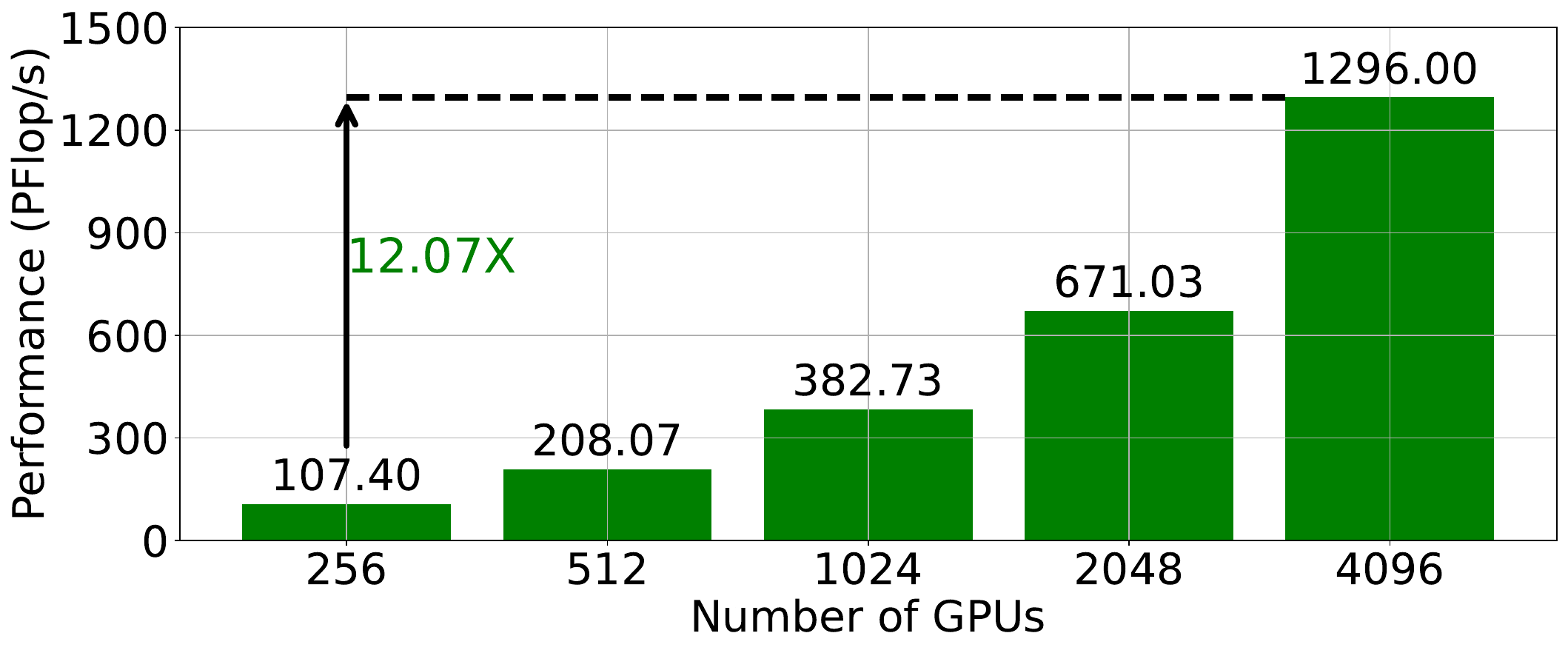}
    \caption{Performance of the \build phase on \alps.}
      \label{fig:tc-distance}    
\end{figure}

We assess the performance of the new TC-based INT8 distance calculation kernels (cast into a SYRK call) used for the \texttt{Build} phase, as described in Section~\ref{subsec:distances}. We run performance scalability up to $1,024$ nodes of \alps. Figure~\ref{fig:tc-distance} displays decent weak performance scalability of the distance calculation kernels when activating INT8 TC, achieving up to $1.296$ mixed-precision ExaOp/s on $4,096$ GH200 Superchips, a 12X speedup compared to $256$ GH200 Superchips (i.e., 75\% parallel efficiency). 
 

\subsection{Performance of the MxP Cholesky (\texttt{Associate}) Across GPU Hardware Generations}
\begin{figure*} [htb!]
  \centering 
    \begin{subfigure}{.3\linewidth}
    \centering 
    \includegraphics[width=\linewidth]{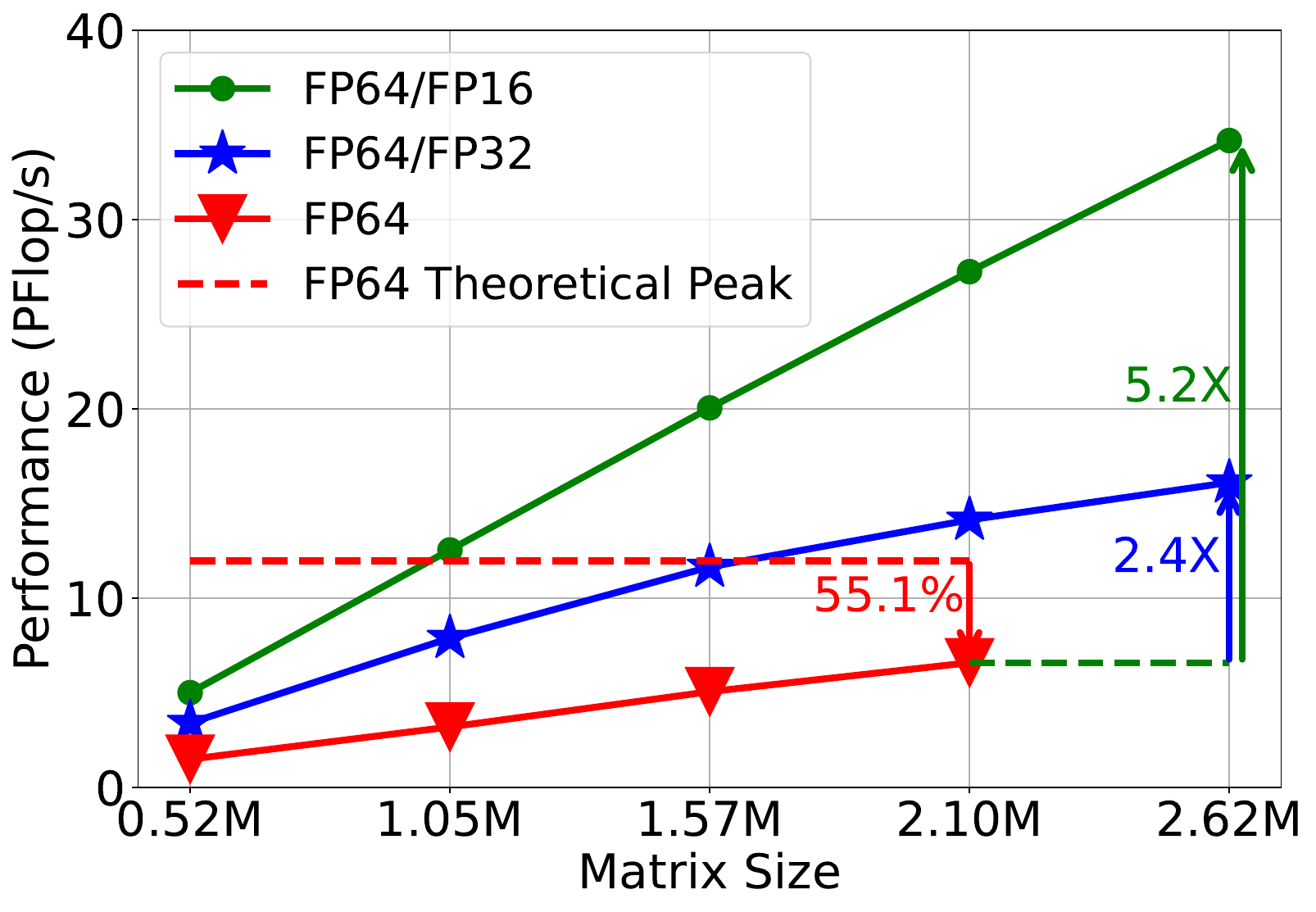}
    \caption{256 nodes (1536 V100 GPUs).}\label{fig:perf-summit-256}
  \end{subfigure}
    \begin{subfigure}{.3\linewidth}
    \centering 
    \includegraphics[width=\linewidth]{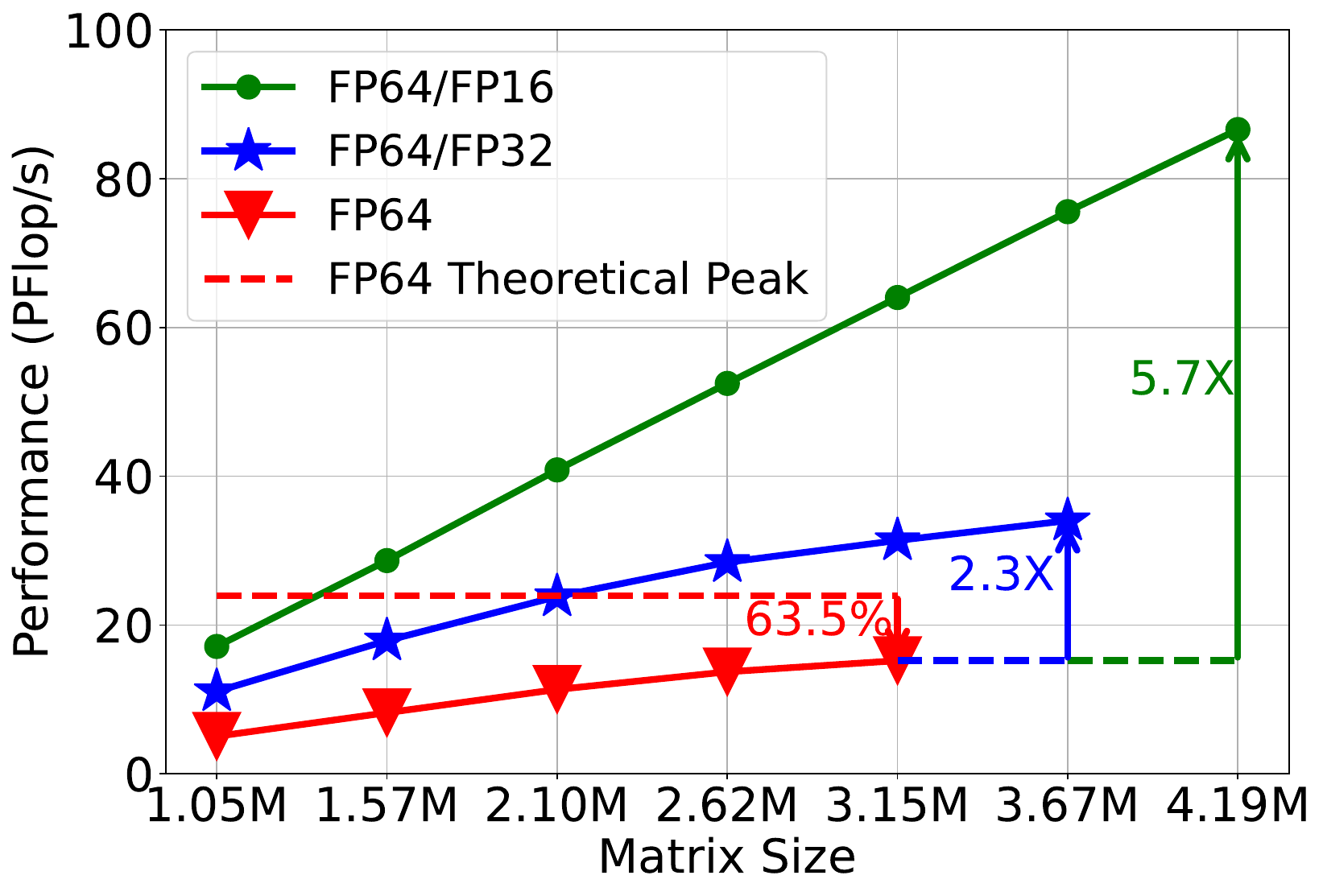}
    \caption{512 nodes (3072 V100 GPUs).}\label{fig:perf-summit-512}
  \end{subfigure}
    \begin{subfigure}{.3\linewidth}
    \centering 
    \includegraphics[width=\linewidth]{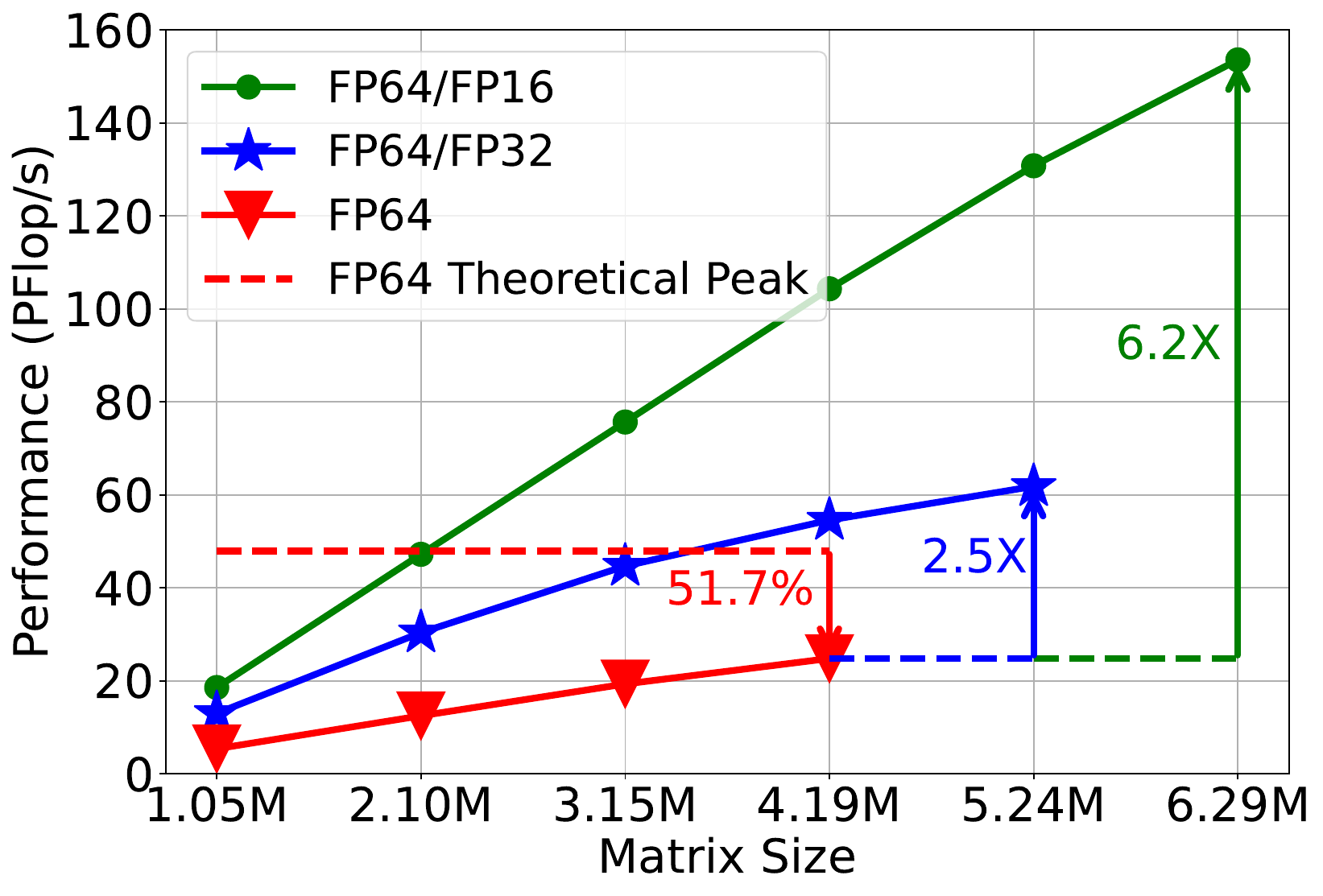}
    \caption{1024 nodes (6144 V100 GPUs).}\label{fig:perf-summit-1024}
  \end{subfigure}
    \caption{Performance scalability of the \texttt{Associate} phase of the KRR-based GWAS ($N_P = N_S$) on \summit.}
\label{fig:assess-summit}
\end{figure*}

\begin{figure*} [htb!]
  \centering 
    \begin{subfigure}{.3\linewidth}
    \centering 
    \includegraphics[width=\linewidth]{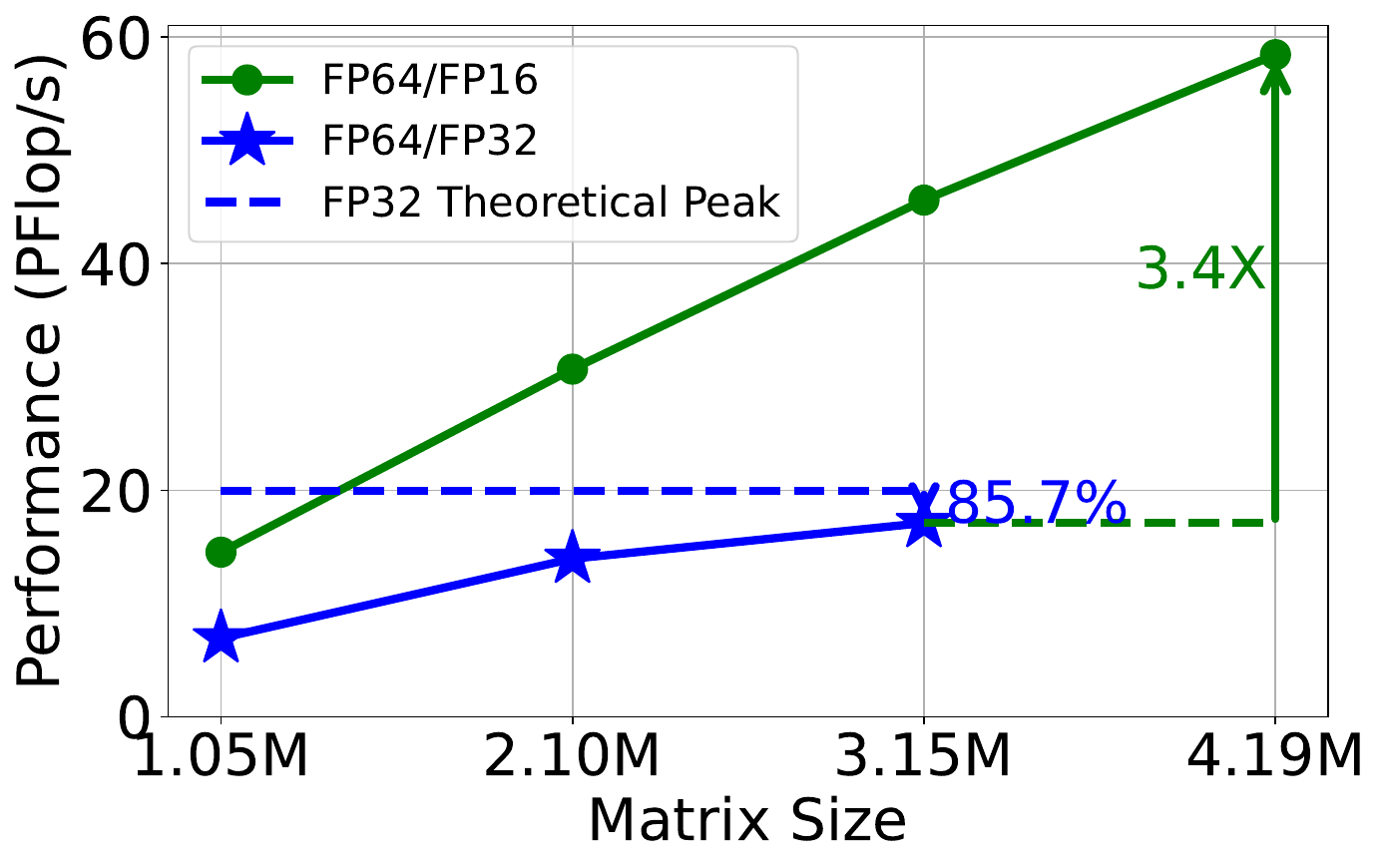}
    \caption{256 nodes (1024 A100 GPUs).}\label{fig:perf-leo-256}
  \end{subfigure}
    \begin{subfigure}{.3\linewidth}
    \centering 
    \includegraphics[width=\linewidth]{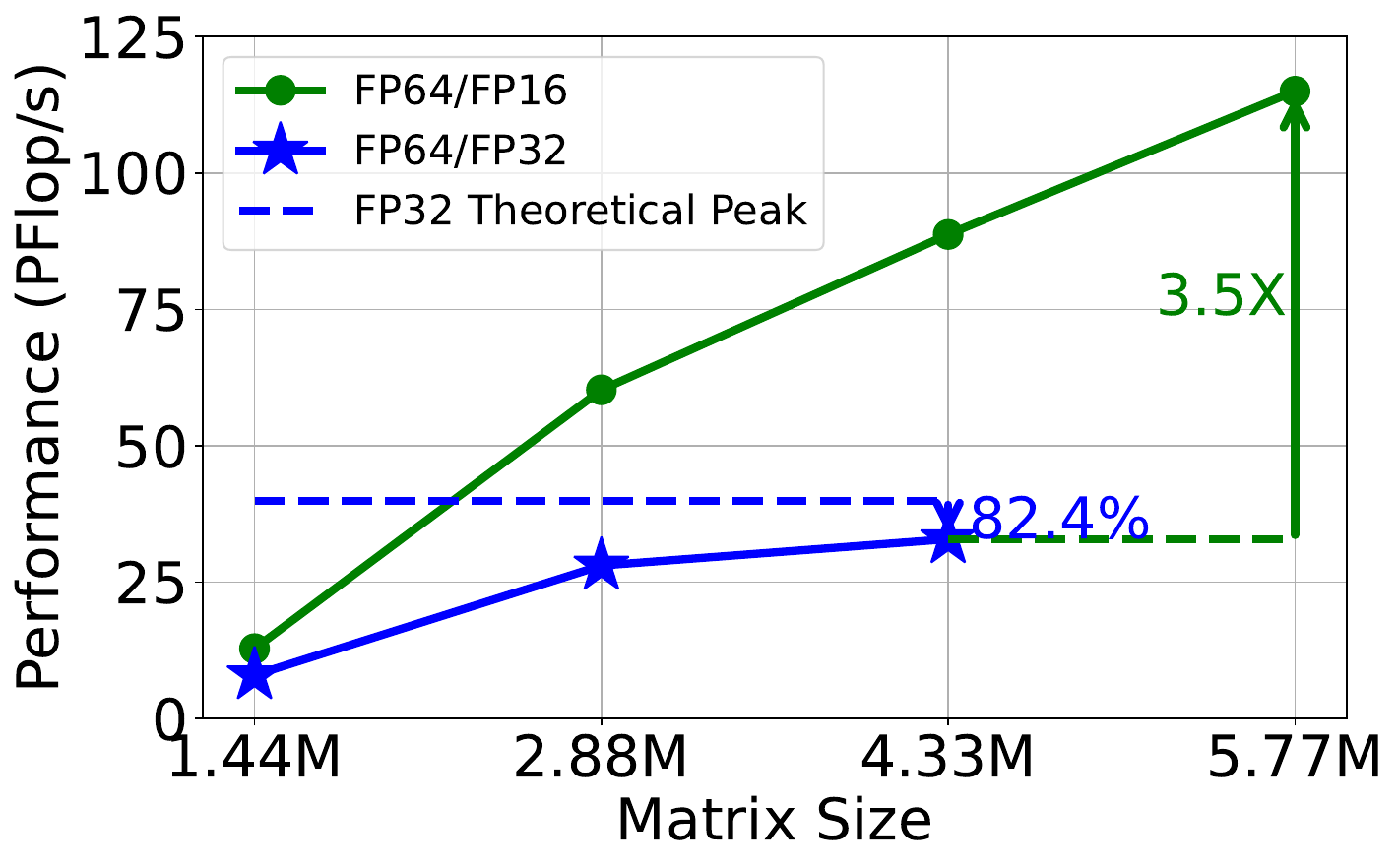}
    \caption{512 nodes (2048 A100 GPUs).}\label{fig:perf-leo-512}
  \end{subfigure}
    \begin{subfigure}{.3\linewidth}
    \centering 
    \includegraphics[width=\linewidth]{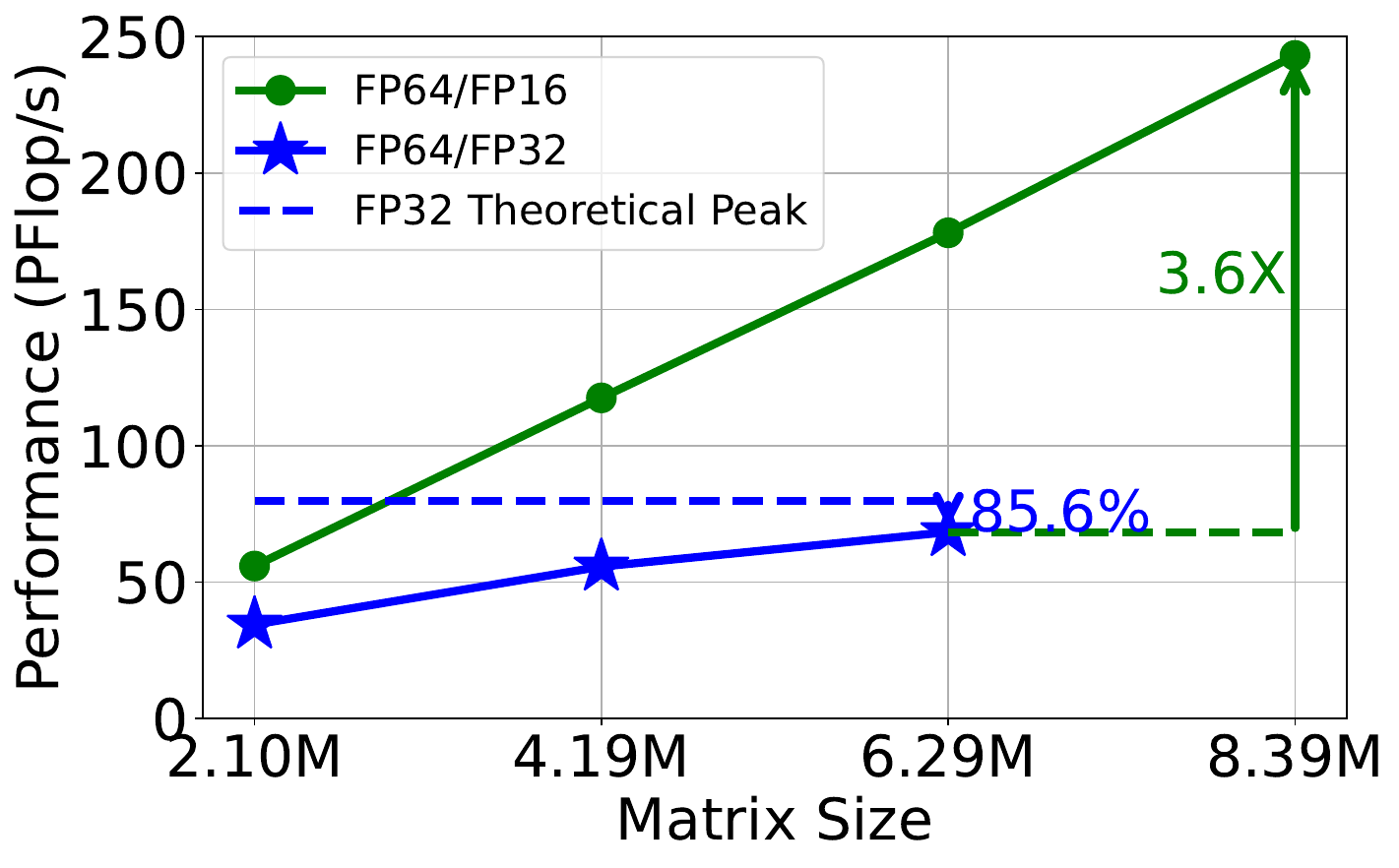}
    \caption{1024 nodes (4096 A100 GPUs).}\label{fig:perf-leo-1024}
  \end{subfigure}
    \caption{Performance scalability of the \texttt{Associate} phase for the KRR-based GWAS ($N_P = N_S$) on \leonardo.}
\label{fig:assess-leo}
\end{figure*}

\begin{figure*} [htb!]
  \centering 
    \begin{subfigure}{.3\linewidth}
    \centering 
    \includegraphics[width=\linewidth]{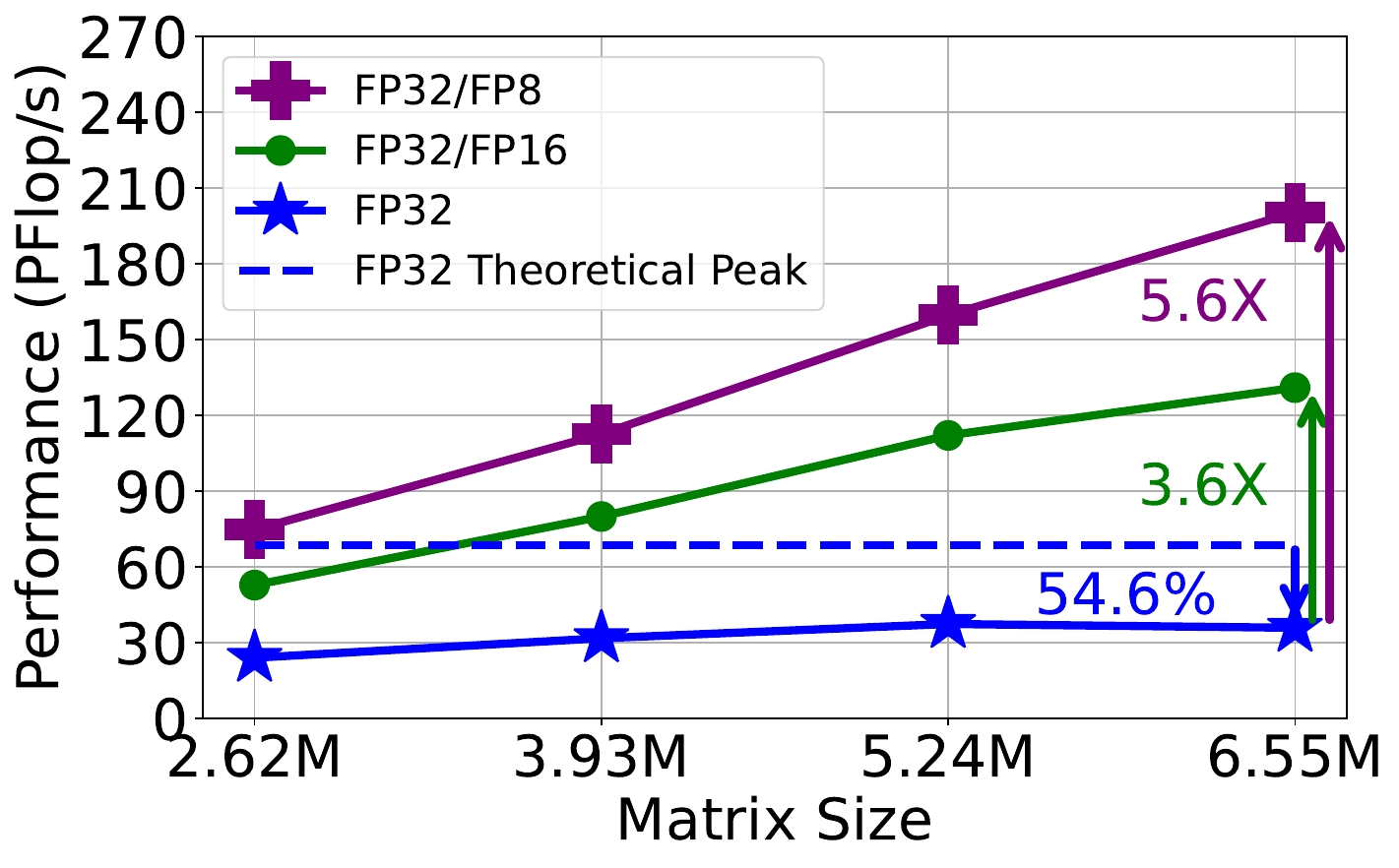}
    \caption{256 nodes (1024 GH200s).}\label{fig:perf-alps-256}
  \end{subfigure}
    \begin{subfigure}{.3\linewidth}
    \centering 
    \includegraphics[width=\linewidth]{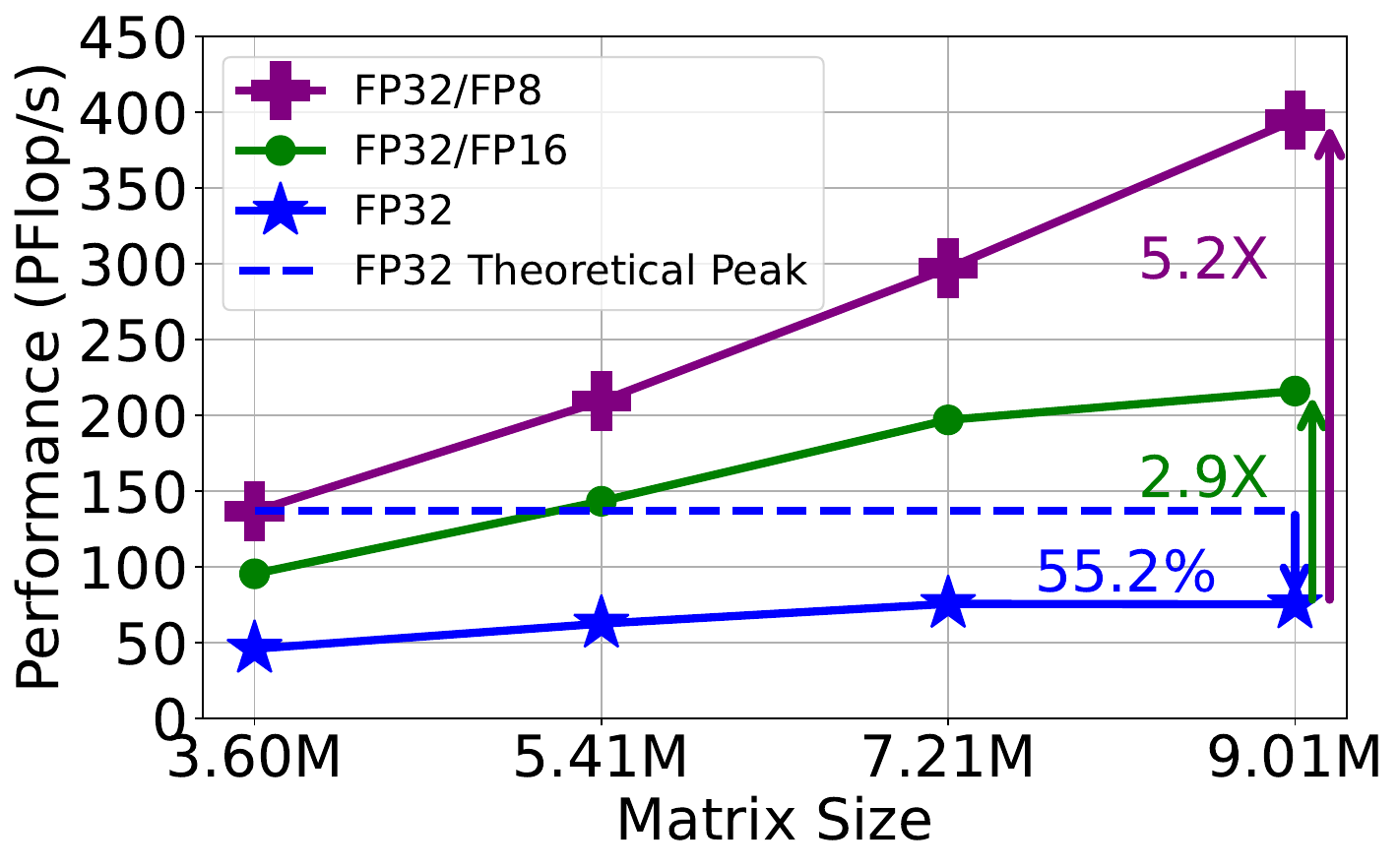}
    \caption{512 nodes (2048 GH200s).}\label{fig:perf-alps-512}
  \end{subfigure}
    \begin{subfigure}{.3\linewidth}
    \centering 
    \includegraphics[width=\linewidth]{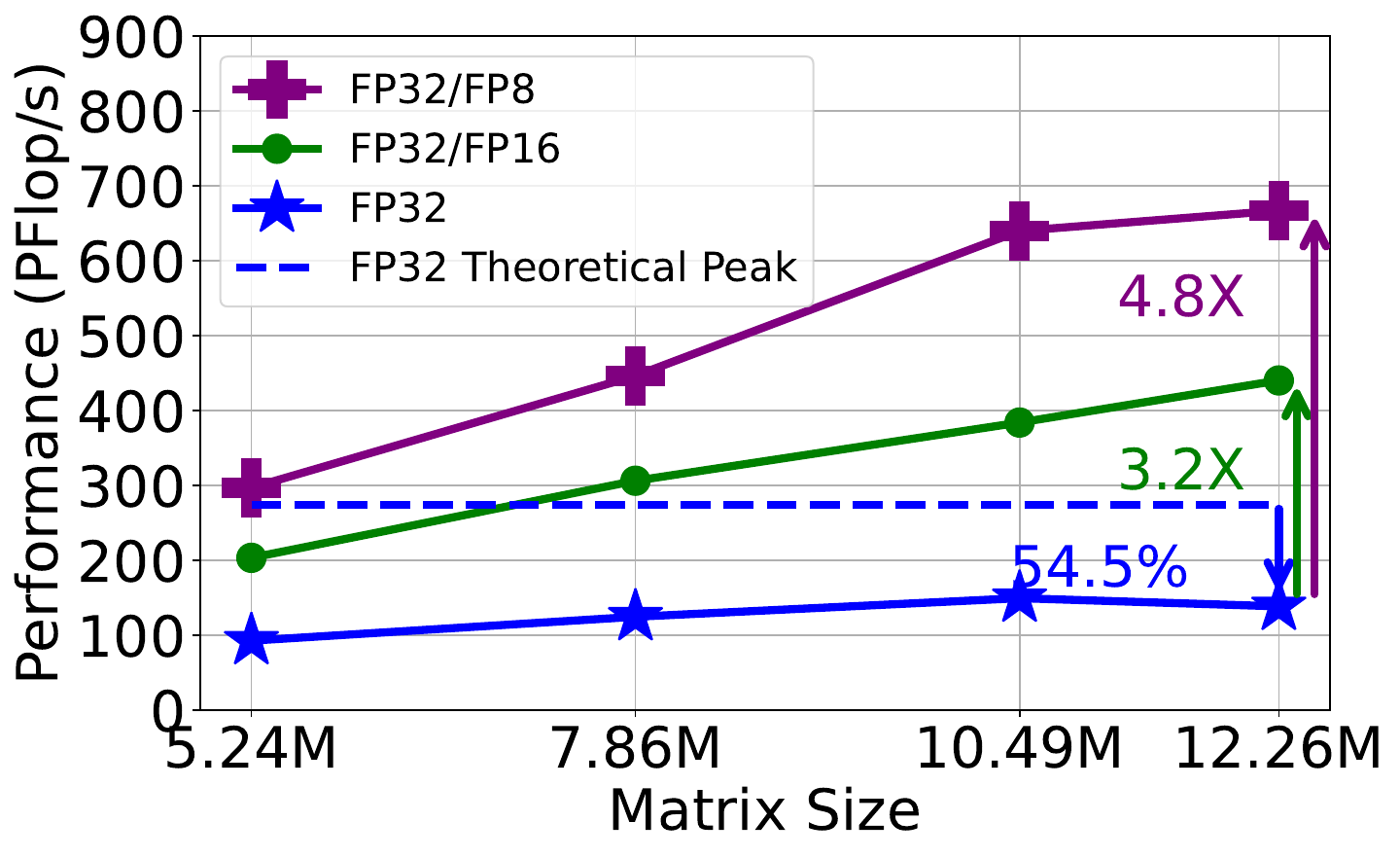}
    \caption{1024 nodes (4096 GH200s).}\label{fig:perf-alps-1024}
  \end{subfigure}
    \caption{Performance scalability of the \texttt{Associate} phase of the KRR-based GWAS ($N_P = N_S$) on \alps.}
\label{fig:assess-alps}
\end{figure*}

Figure~\ref{fig:assess-summit} shows the performance of the \texttt{Associate} phase, which is driven by the MxP Cholesky factorization using FP64/FP32 or FP64/FP16 on various numbers of \summit nodes. On $1024$ nodes, we reach around $62$ PFlop/s and $154$ PFlop/s, which are, respectively, $2.5$X and $6.2$X faster than doing all computations in FP64. 
Figure~\ref{fig:assess-leo} depicts the same performance configurations but using \leonardo nodes with A100 GPUs.
On $1024$ nodes, we reach around $243$ PFlop/s, which is $3.6$X faster than doing all computations in FP32 (or FP64, since FP64/FP32 sustain the same execution rate on A100 GPUs).
Figure~\ref{fig:assess-alps} illustrates again the same performance configurations but with \alps nodes powered by GH200 Superchips.
On $1024$ nodes, we reach around $440$ and $667$ PFlop/s for FP32/FP16 and FP32/FP8, which are, respectively, $3.2$X and $4.8$X faster than doing all computations in FP32.

An interesting trend for the three aforementioned figures with different GPU hardware generations is the widening gap between communication/computation, which further challenges parallel efficiency in presence of low precisions, as we increase the number of nodes for each system. As the low precision arithmetic throughput gets a boost by a factor of $3$X between GPU generations, the cost of data movement increasingly impedes strong scalability. 

\begin{figure}[t]
\center
\begin{subfigure}{.472\linewidth}
\includegraphics[width=1\linewidth]{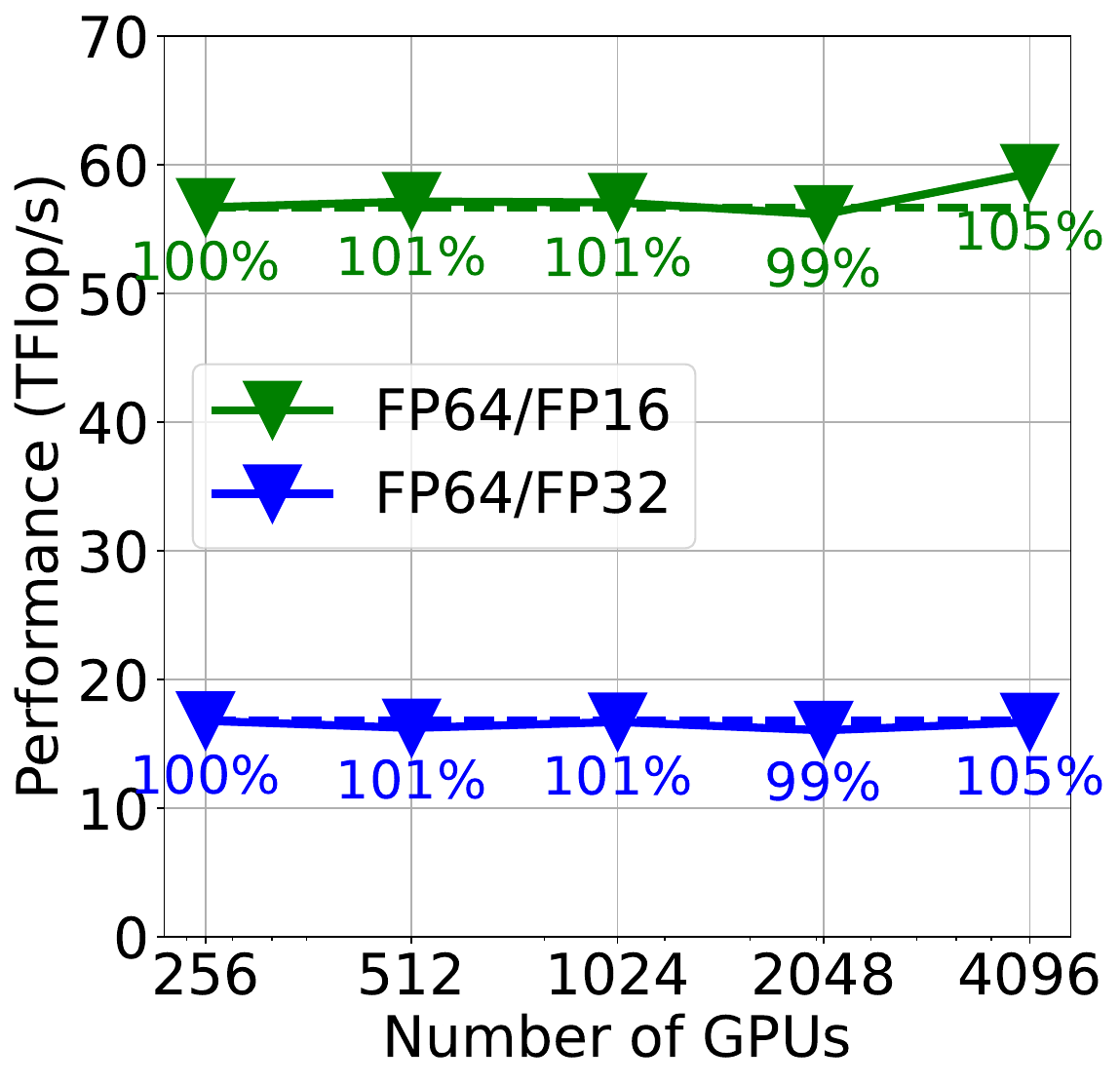}
\caption{Weak scalability.}
\label{fig:perf-leonardo_weak}
\end{subfigure}
\begin{subfigure}{.472\linewidth}
\includegraphics[width=1\linewidth]{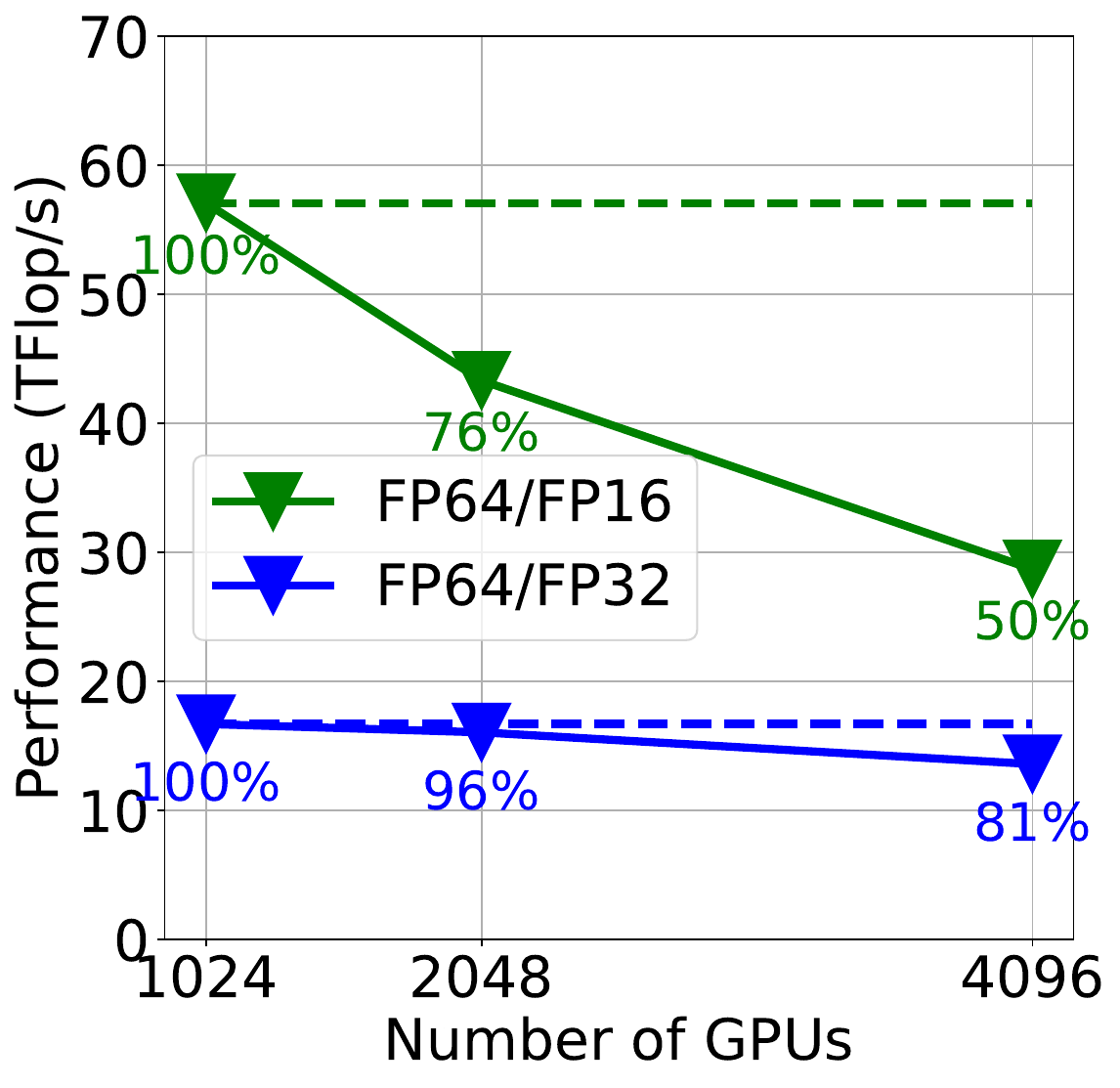}
\caption{Strong scalability.}
\label{fig:perf-leonardo_strong}
\end{subfigure}
\caption{\texttt{Associate} on \leonardo normalized per GPU.}
\end{figure}
\begin{figure}[t]
\center
\begin{subfigure}{.49\linewidth}
\includegraphics[width=1\linewidth]{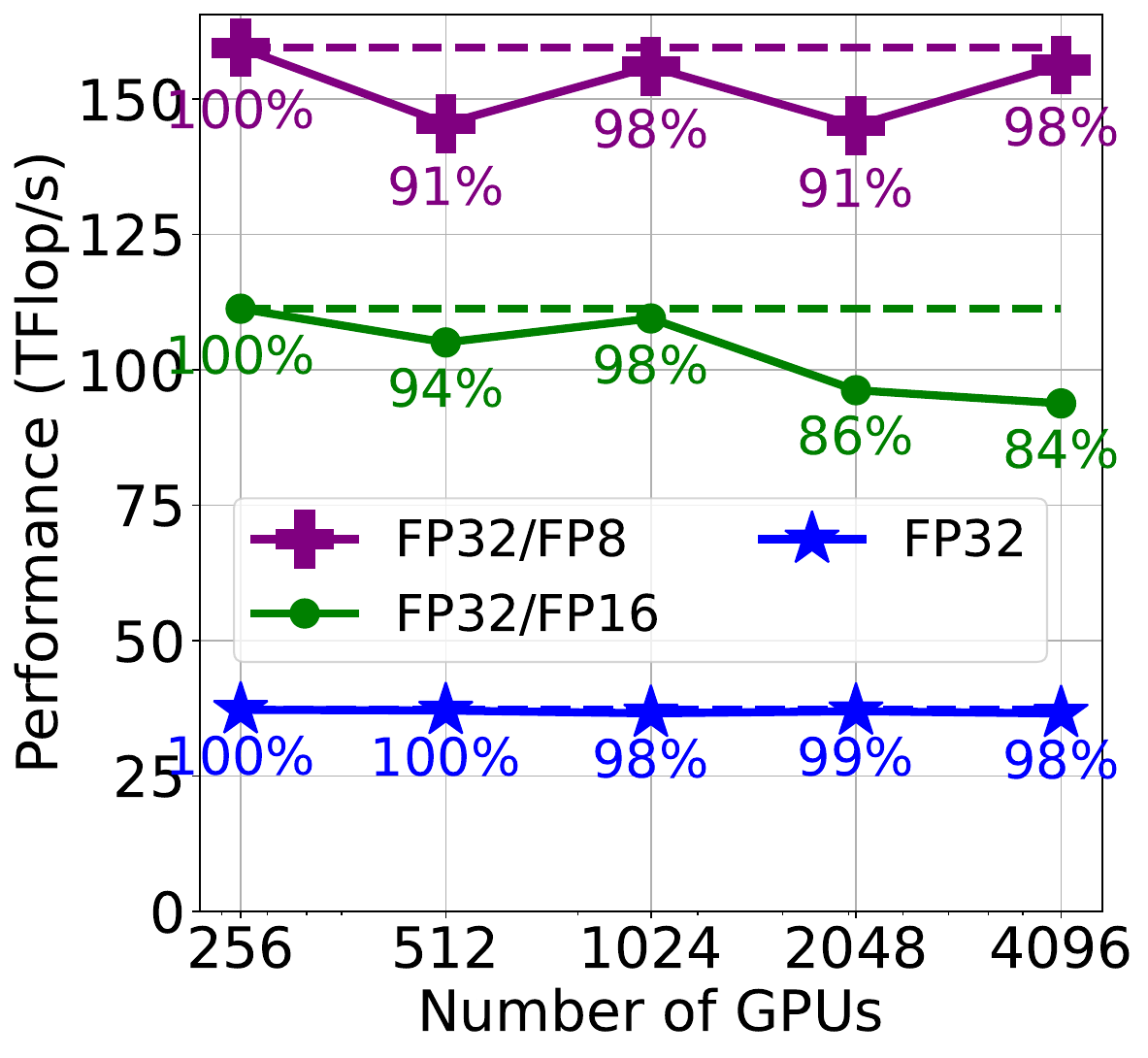}
\caption{Weak scalability.}
\label{fig:perf-alps_weak}
\end{subfigure}
\begin{subfigure}{.49\linewidth}
\includegraphics[width=1\linewidth]{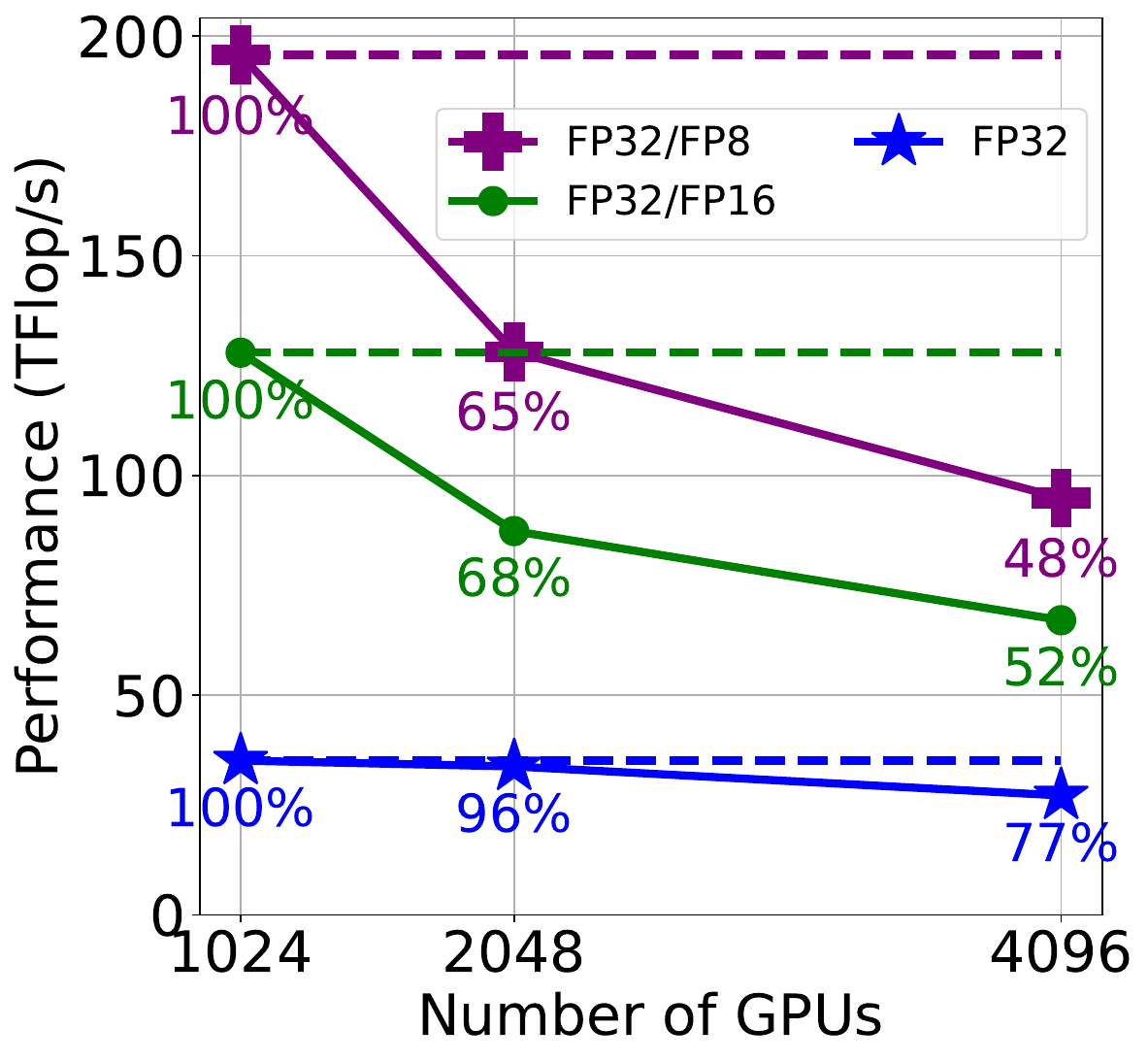}
\caption{Strong scalability.}
\label{fig:perf-alps_strong}
\end{subfigure}
\caption{\texttt{Associate} on \alps normalized per GPU.}
\end{figure}

While Figures~\ref{fig:perf-leonardo_weak} and~\ref{fig:perf-alps_weak} show the near perfect weak scaling on \leonardo and \alps of the mixed-precision \texttt{Associate} phase with around $57$ and $159$ TFlop/s per GPU, respectively, Figures~\ref{fig:perf-leonardo_strong} and~\ref{fig:perf-alps_strong} displays the performance impact of strong scaling on parallel efficiency using \leonardo and \alps, respectively. The parallel efficiency drops to around~50\% on 4096 GPUs when engaging lower precisions (FP16/FP8) during the \texttt{Associate} phase as opposed to~81\% and 77\% for higher precisions (FP64/FP32). Indeed, the workloads per GPU decrease even faster when FP16/FP8 is involved due to lower hardware occupancy, which exposes the data movement bottleneck. Nevertheless, there remains an advantage in engaging low precision hardware features not only in terms of performance but also in terms of memory footprint. Leveraging CUDA-aware MPI and tuning lookahead optimizations can help overcome this bottleneck. 


\subsection{Performance of KRR-based Multivariate GWAS on \alps}
\begin{figure}[t]
\center
\begin{subfigure}{.49\linewidth}
\includegraphics[width=1\linewidth]{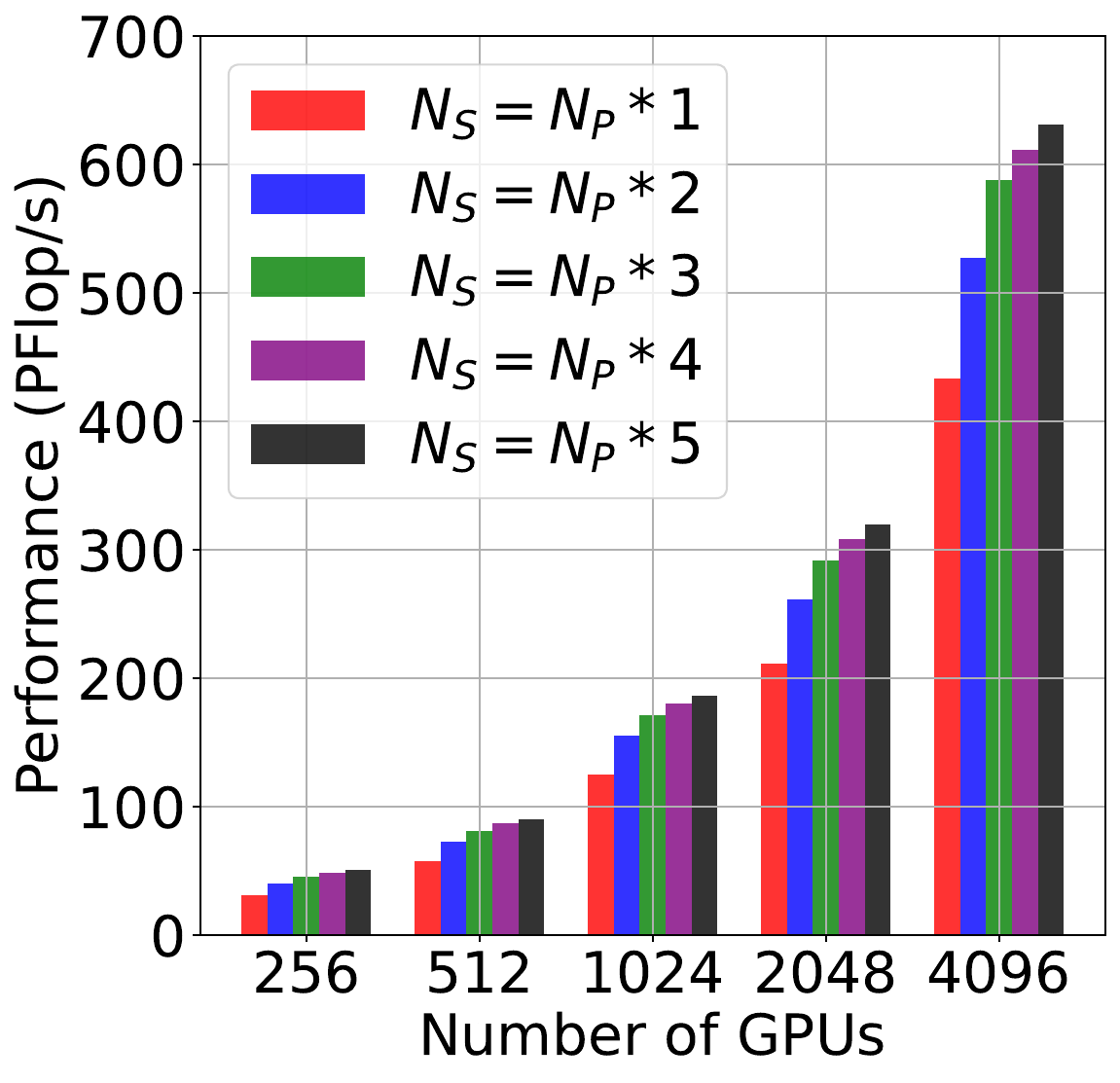}
\label{fig:perf-alps_krr_snp_hp}
\end{subfigure}
\begin{subfigure}{.49\linewidth}
\includegraphics[width=1\linewidth]{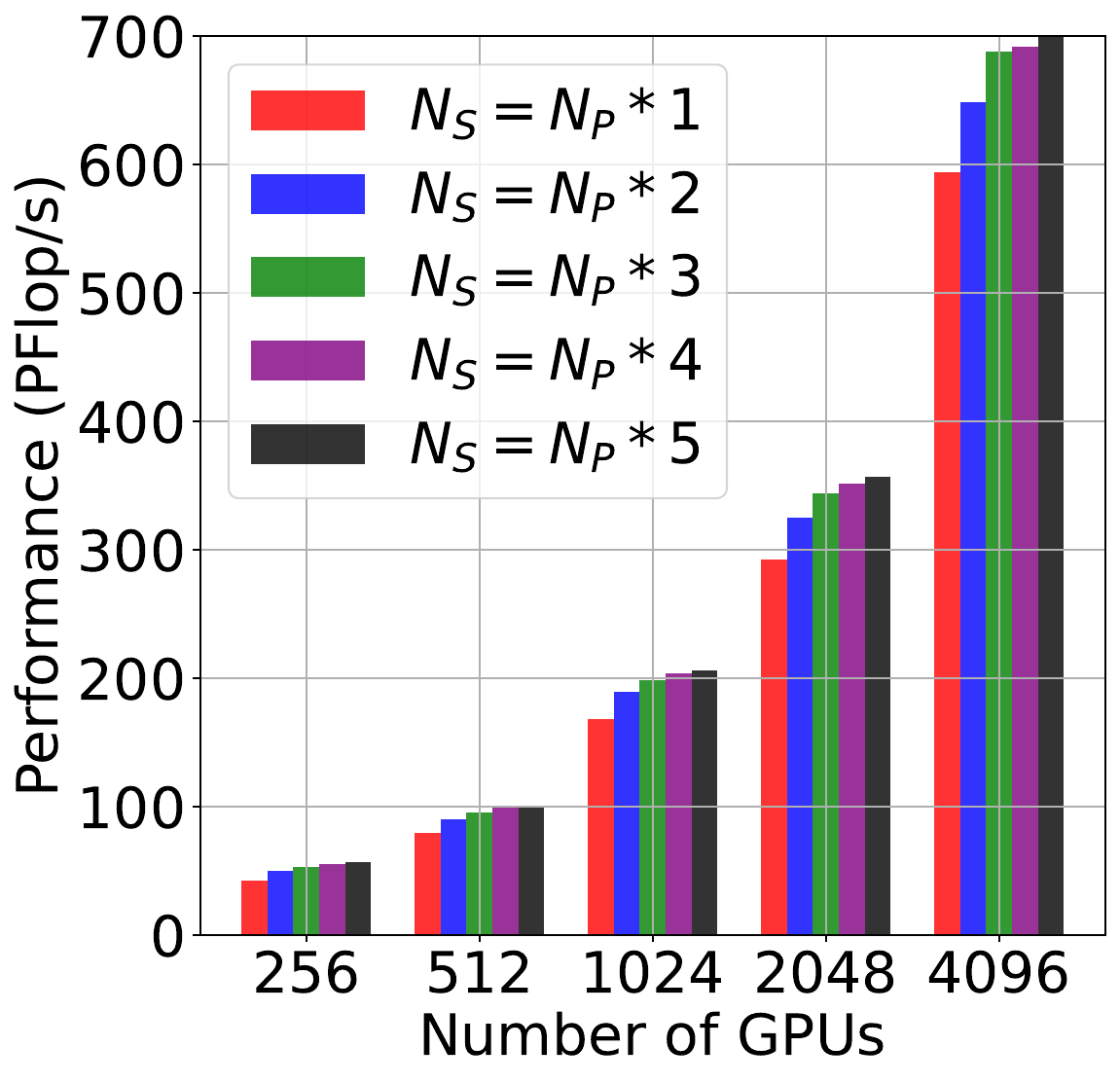}
\label{fig:perf-alps_krr_snp_fp8}
\end{subfigure}
\caption{{FP32/FP16 (left) and FP32/FP8 (right) weak scaling of KRR-based GWAS on \alps.}}
\label{fig:perf-alps_krr_snp}
\end{figure}

Figure~\ref{fig:perf-alps_krr_snp} shows decent weak scalability achieved by the MxP KRR-based Multivariate GWAS (i.e., FP32/FP16 and FP32/FP8 configurations), combining both \texttt{Build} and \texttt{Associate} phases, maxing out the device memory, up to $4,096$ GPUs on \alps. As we increase $N_S$, the throughput of the dominating \texttt{Build} phase increases (see the algorithmic complexity in Section~\ref{subsec:complexity}), which further accelerates the overall KRR-based multivariate GWAS. Since lowering precision from FP16 to FP8 benefits only the \texttt{Associate} phase, the overall performance improvement decreases between MxP FP16/FP8 as $N_S$ increases. 

\subsection{Large-Scale Experiments on \alps and System Performance Comparisons}

\begin{figure*} [htb!]
  \centering 
    \begin{subfigure}{.195\linewidth}
    \centering 
    \includegraphics[width=\linewidth]{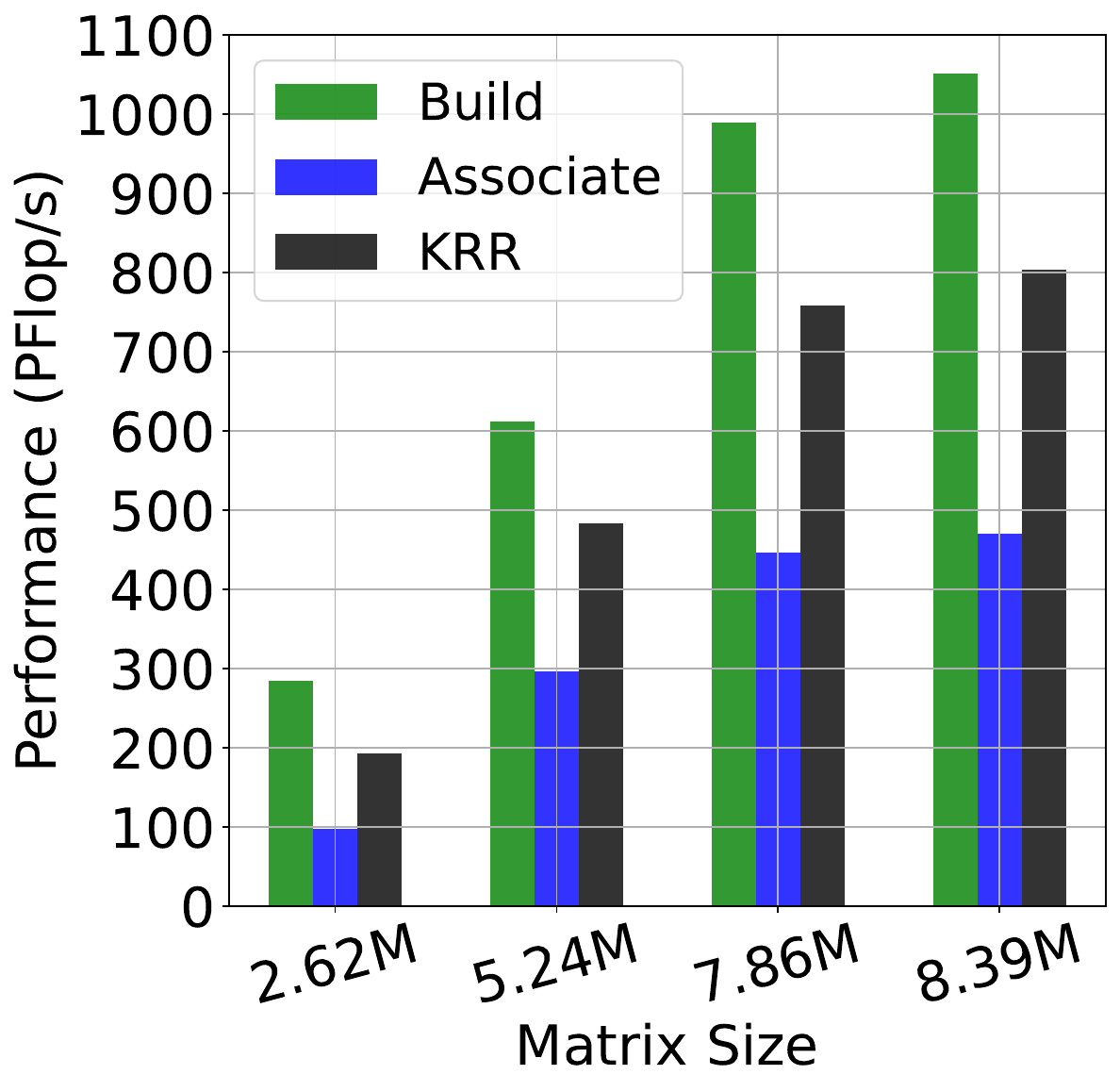}
    \caption{1024 nodes on \alps.}\label{fig:krr-perf-alps-1024}
  \end{subfigure}
    \begin{subfigure}{.195\linewidth}
    \centering 
    \includegraphics[width=\linewidth]{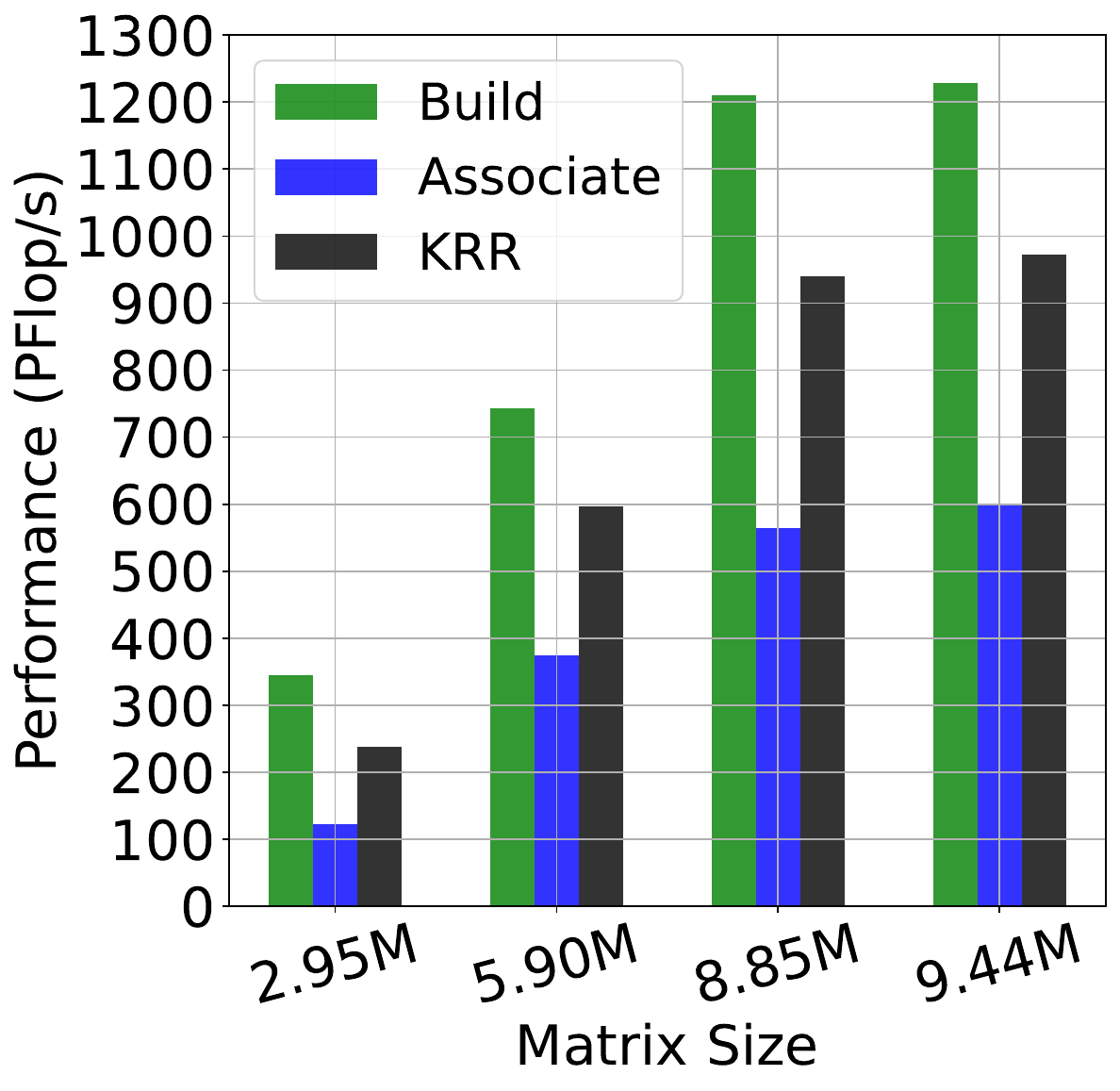}
    \caption{1296 nodes on \alps.}\label{fig:krr-perf-alps-1296}
  \end{subfigure}
    \begin{subfigure}{.195\linewidth}
    \centering 
    \includegraphics[width=\linewidth]{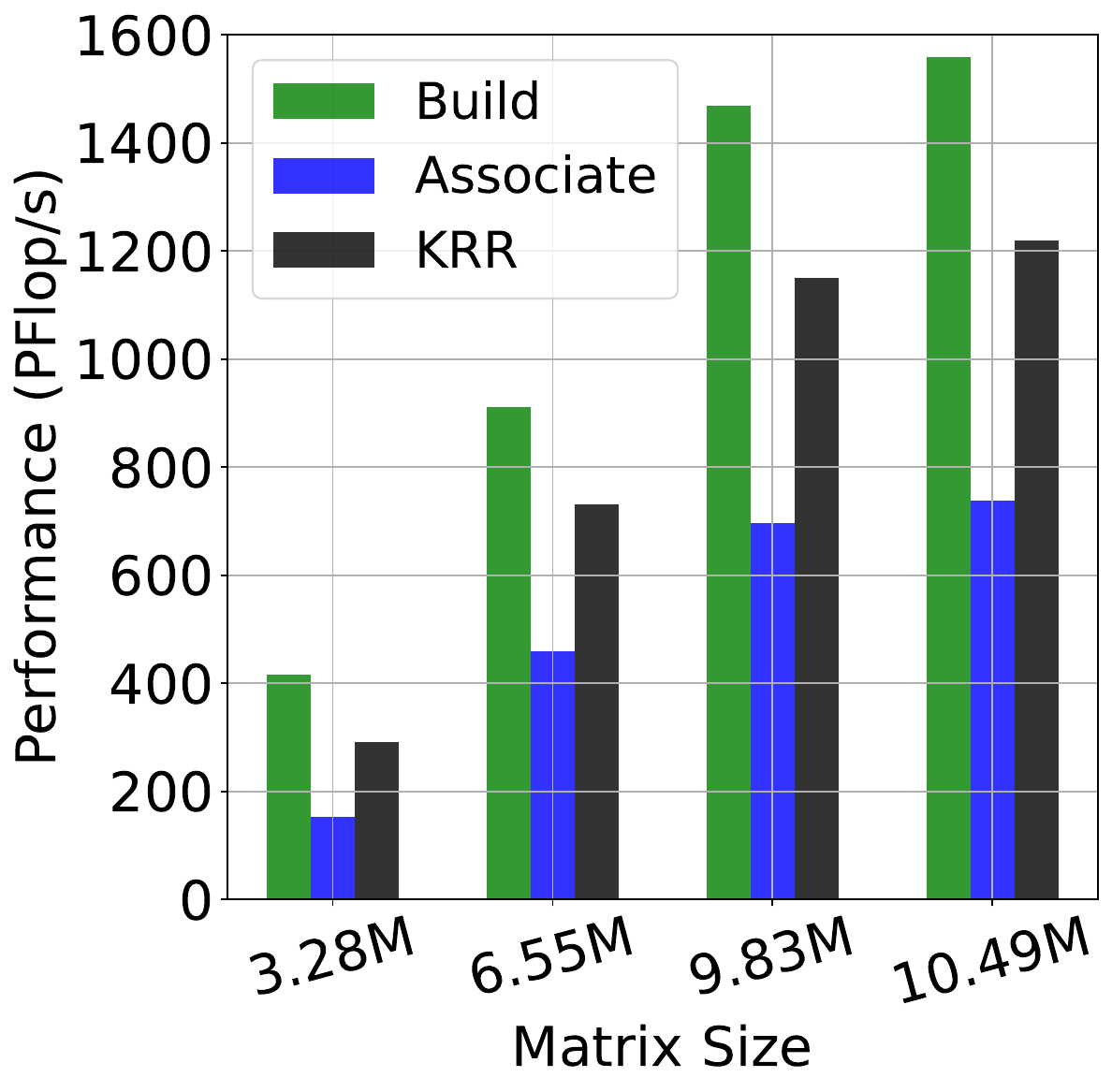}
    \caption{1600 nodes on \alps.}\label{fig:krr-perf-alps-1600}
  \end{subfigure}
      \begin{subfigure}{.195\linewidth}
    \centering 
    \includegraphics[width=\linewidth]{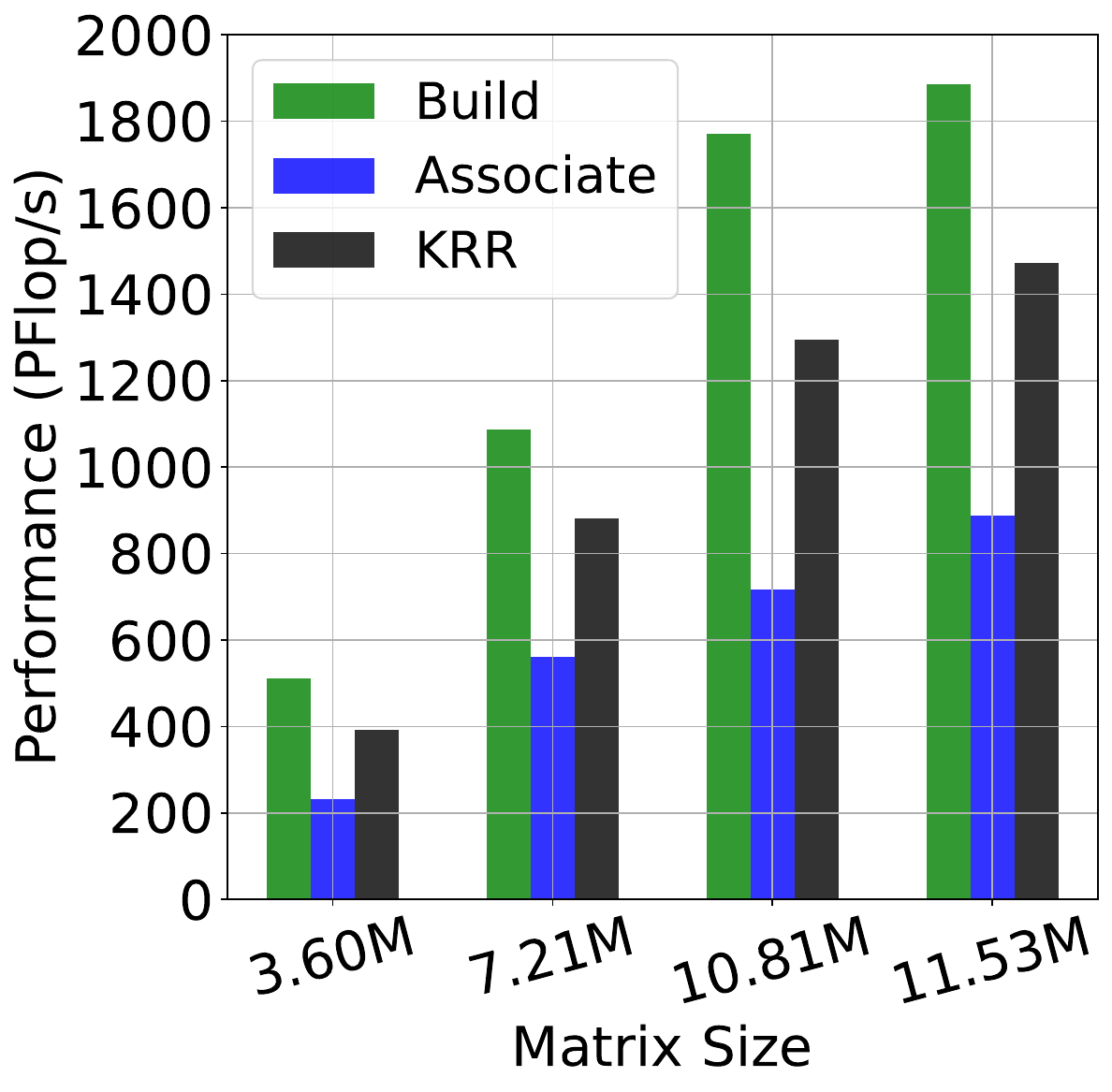}
    \caption{1936 nodes on \alps.}\label{fig:krr-perf-alps-1936}
  \end{subfigure}
      \begin{subfigure}{.195\linewidth}
    \centering 
    \includegraphics[width=\linewidth]{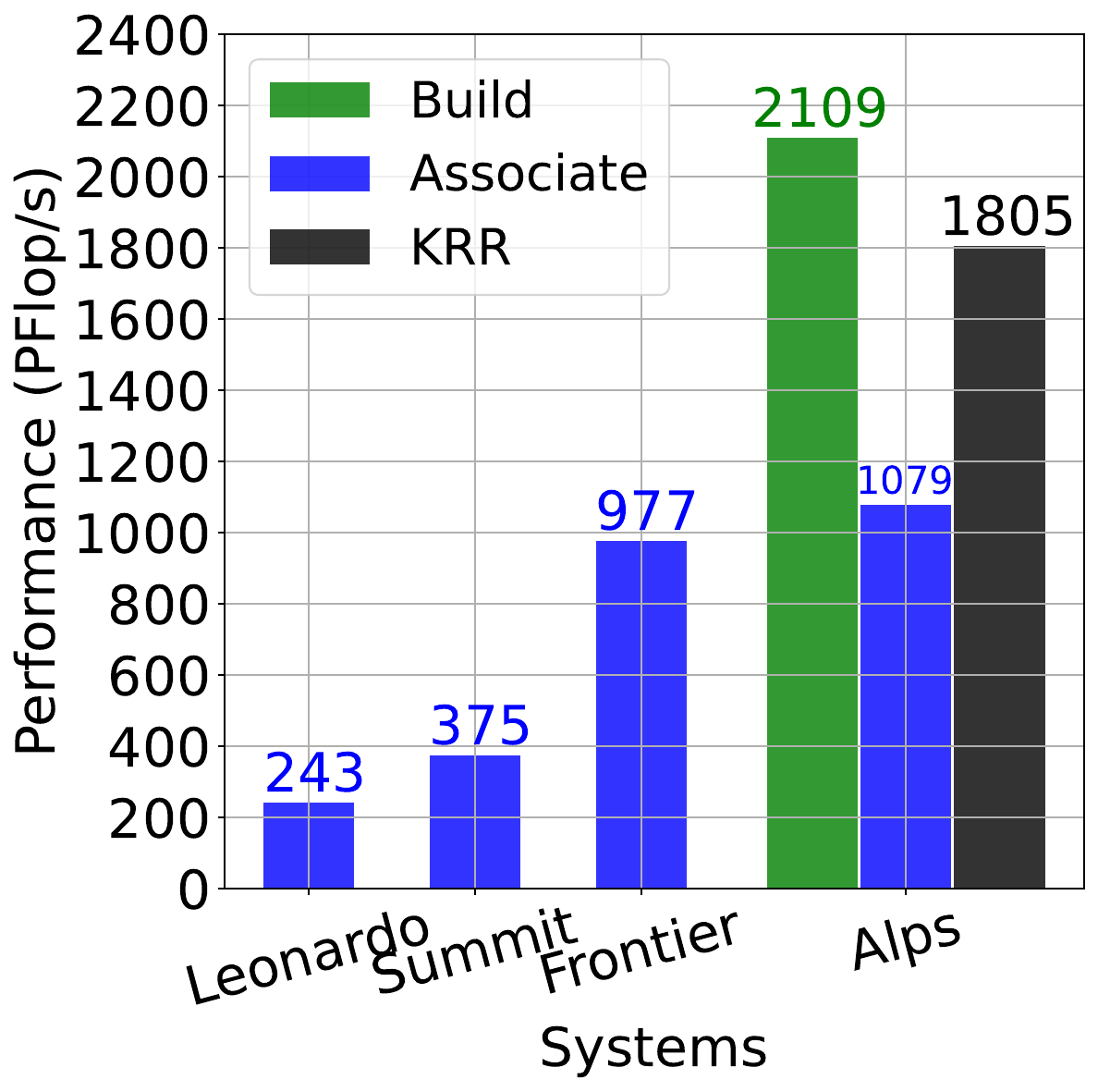}
    \caption{Across systems.}\label{fig:perf-max}
  \end{subfigure}
    \caption{Large-scale KRR-based multivariate GWAS. In (a)-(d), $N_P = N_S$; $N_P < N_S$ in (e).}
\label{fig:krr-alps}
\end{figure*}

Figures~\ref{fig:krr-alps}a-d present the performance breakdown of large-scale KRR-based multivariate GWAS using various matrix sizes on $1,024$, $1,296$, $1,600$, and $1,936$ nodes of \alps. The \texttt{Build} phase scores the highest throughput and enables to maintain weak scalability as the problem sizes and the number of nodes increase. Figure~\ref{fig:krr-alps}e illustrates the absolute performance comparison of the \texttt{Associate} phase using problem sizes that occupy most of the GPU memory across four different systems: $4,096$ GPUs (1/3$^{rd}$) of \leonardo, $18,432$ GPUs (2/3$^{rds}$) of \summit, $36,100$ GPUs (nearly full) of \frontier, and $8,100$ GPUs (4/5$^{ths}$) of \alps. The \texttt{Associate} phase on \alps scores speedups of up to 2X over \leonardo using the same number of GPUs (4X with twice the number of GPUs) and 3X against \summit with less than a half of the number of GPUs. These speedups highlight the hardware technology scaling across NVIDIA GPU hardware generations. Moreover, this phase on \alps achieves slightly higher performance than \frontier with less than one fourth of the number of GPUs. Looking at the large-scale run on \alps, using 13M patients and 20M SNPs, we achieve $2.109$ mixed-precision ExaOp/s (over all precision types) for the \texttt{Build} phase, which boosts the performance of the overall KRR-based multivariate GWAS to $1.805$ mixed-precision ExaOp/s. 

We conclude with a comparison of the expanded headroom in performance offered by our distributed-memory mixed-precision KRR solver against the state-of-the-art highly efficient REGENIE~\cite{Mbatchou2021} shared-memory software for whole-genome regression for CPUs, described in Section IV.  Giving the BLAS3-intensive stacked ridge regression REGENIE implementation credit for full theoretical peak on a state-of-the-art CPU, e.g., the 
dual-socket 96-core 2.40 GHz AMD Genoa 9654 of KAUST's HPE-Cray \shaheen-3 at $7.372$ TFlop/s, and using our best actual performance of $1.805$ mixed precision ExaOp/s on 4/5$^{ths}$ of the \alps system, the performance ratio is about five orders of magnitude.

\section{IMPLICATIONS }
\label{sec:implications}
\ifthenelse {\boolean{GuideOn}} {\textbf{implications for future systems and applications (1 p max)\\
}} 

Pushing the genotypic ($N_S$) and population ($N_P$) dimensions of KRR and other nonlinear kernel methods for GWAS to the extremes of this paper issues a loud welcome to genome scientists to exploit the mixed precision exascale computing environment.  Datasets like FinnGen~\cite{Kurki2023}, currently at half a million Finns (about 10\% of the population), can be extended and analyzed with GWAS at the national scale. Extending patient populations to 13 million, as herein synthesized, democratizes GWAS, accommodating the full population of 63\% of the world's countries.  Many countries' populations are underrepresented in medical studies and therefore underserved in any medicine that incorporates their particular SNP and confounder characteristics.  
It is also interesting (and straightforward) to expand the input variables to include diverse environmental factors (eGWAS)~\cite{Virolainen2022}).
Once a population's $G$ matrix is transformed into a KRR $K$ matrix, its patient data is effectively anonymized for unrestricted analysis, which should lead to progress on broader academic fronts.

The benefits of GWAS reach far beyond human health. GWAS has become foundational to the study of grains, for example, with the aim of feeding 10 billion humans through greater disease resistance and less environmental impact through breeding to confer salt and drought tolerance that relieves pressure on fresh water supplies.
Campaigns to this end on wheat, rice, quinoa, and barley are ongoing on KAUST's supercomputers.  The wheat genome is about six times larger than the human genome and exhibits vast variation after millennia of regional domestication.

Since the first study in 2005 on 96 patients~\cite{Klein2005GWAS}, GWAS has been used to identify more than $500$K SNP loci that are associated with various complex human diseases or traits. However, these are statistical associations that do not of themselves reveal mechanisms. Current efforts underway to integrate GWAS with functional genomics for understanding of gene regulation to accompany the associations multiply the potential benefits of routine, low-cost GWAS throughout functional biology~\cite{li_ritchie2021}.

On an algorithmic note, the mixed precision techniques of this paper, which exploit the wide range of norm variation in $K$ matrix tiles, offer a good measure of memory footprint reduction, but additional and potentially even greater data sparsity may be available from exploiting the smoothness of matrix tiles in the form of low-rank replacements of dense tiles in their algebraic manipulation.  It has been shown, for instance, that KRR can exploit low rankness in the Hierarchically Semi-separable (HSS) sense~\cite{chavez2020}. A previous Gordon Bell finalist demonstrated on CPUs only how mixed precision and low rank structure can be exploited {\em simultaneously} within the same PaRSEC dynamic runtime system employed in this work in the context of covariance matrices from spatial statistics~\cite{caoGB2022}. Further GWAS exploration of this synergy lies ahead.


It is also noteworthy that the increasing availability of data on the 3D organization of genomes has spurred the development of 3D GWAS. This approach aims to identify associations between disease or trait susceptibility and a wider range of genetic variants, particularly non-coding SNPs, which traditional 2D GWAS often misses. The 3D GWAS approach achieves this by capturing interactions between: (1) Distal regulatory elements and genes, i.e.,  elements located further away in the genome that can influence gene expression but are not captured by 2D proximity analysis, and (2) Co-expressed or co-suppressed genes, i.e., genes in close spatial proximity within the 3D genome structure, even if distant in the linear sequence that can exhibit coordinated activity. This spatial co-localization might influence their combined effect on traits.
3D GWAS has the potential to capture aspects of both LD and epistatic interactions, offering a more comprehensive view of the genetic basis of complex traits.
Our algorithmic solution can leverage these 3D genomic contact maps and apply spatial ordering techniques to further expose data sparsity to maximize performance. Furthermore, our approach embraces the increasing complexity of omics data. As additional data layers, such as epigenomic information, become available for larger cohorts, they can be incorporated to further assist with matrix ordering. This can potentially decrease the computational burden, enhance qualitative prediction, and facilitate analysis of ever-growing datasets.

From a hardware perspective, the pace of GPU innovations driven by the AI market, creates exciting opportunities for KRR-based multivariate GWAS, an approach that is overlooked due to its inefficient workloads (i.e., similarity comparisons) and the high computational complexity. In this paper, using distance kernels rather than similarity kernels and using MxP Cholesky, we address both fronts thanks to low-precision TC hardware features of NVIDIA \texttt{Ampere} and \texttt{Hopper} GPU accelerators. The next generation of NVIDIA Superchips, codenamed \texttt{Blackwell}, is expected to deliver more than twice the throughput of \texttt{Hopper} for each INT8/FP16/FP8 precision, while introducing a new FP4 format reaching 40 PFlop/s theoretical peak performance. This unprecedented energy-efficient, low-precision throughput may boost the performance of current mixed-precision linear algebra that supports GWAS workloads and beyond~\cite{ltaief-reckless}.

\section*{ACKNOWLEDGMENTS}
\small{
For computer time, this research used \shaheen-3 and \ibex at the Supercomputing Laboratory of the King Abdullah University of Science and Technology (KAUST) in Thuwal, Saudi Arabia; \frontier and \summit at the Oak Ridge Leadership Computing Facility at the US DOE's Oak Ridge National Laboratory; \leonardo at CINECA in Bologna, Italy; and \alps at CSCS in Lugano, Switzerland. The authors are deeply grateful to the OLCF, CINECA, and CSCS for the allocations that allowed scaling runs after completing the science runs at KAUST and to NVIDIA for brokering the CSCS and CINECA connections. We also thank Bilel Hadri of the KSL team at KAUST,  Aurelien Bouteiller and Thomas Herault from ICL of the University of Tennessee, Gabriele Paciucci, George Bosilca, and David Ruau of NVIDIA, and Maria Grazia Guiffreda of CSCS for their essential support in facilitating this work. Rabab Alomairy was supported on an Ibn Rushd post-doctoral fellowship from KAUST.
}



\bibliographystyle{unsrt2authabbrvpp}
\bibliography{ref}

\end{document}